\documentclass[aps,prd,onecolumn,groupedaddress,floatfix,longbibliography,superscriptaddress,notitlepage]{revtex4-2}

\usepackage{amsmath}
\usepackage{amssymb}
\usepackage{amsfonts}
\usepackage{bbold}
\usepackage{bm}
\usepackage{cancel}
\usepackage{times,float}
\usepackage{graphicx}
\usepackage[usenames,dvipsnames,svgnames]{xcolor}
\usepackage{hyperref}
\hypersetup{colorlinks=true, linkcolor=NavyBlue, citecolor=PineGreen,urlcolor=cyan}
\usepackage{multirow}
\usepackage{physics}
\usepackage{esint}
\usepackage{caption}
\usepackage{subcaption}
\usepackage{tikz}
\usepackage{pgfplots}
\usepackage{pgfplotstable}
\pgfplotsset{compat = newest}
\usetikzlibrary{positioning, arrows.meta}
\usepgfplotslibrary{fillbetween}
\usepackage{svg}

\begin{document}

\title{Linear and nonlinear transport responses of topological nodal-line semimetals}

\author{L. E. Sosa-Arias}
\email{sosa.luis@ciencias.unam.mx}
\affiliation{Instituto de Ciencias Nucleares, Universidad Nacional Aut\'{o}noma de M\'{e}xico, 04510 Ciudad de M\'{e}xico, M\'{e}xico}

\author{A. Mart\'{i}n-Ruiz}
\email{alberto.martin@nucleares.unam.mx}
\affiliation{Instituto de Ciencias Nucleares, Universidad Nacional Aut\'{o}noma de M\'{e}xico, 04510 Ciudad de M\'{e}xico, M\'{e}xico}

\begin{abstract}
Topological nodal-line semimetals are three-dimensional quantum materials characterized by band crossings that form closed loops in momentum space. In $\mathcal{PT}$-symmetric realizations, these nodal rings are stabilized in the absence of spin-orbit coupling, giving rise to drumhead surface states and unconventional transport responses. In this work, we study charge transport across a nodal-line semimetal containing a finite electrostatic barrier, with both leads described by the same equilibrium material. By solving the corresponding scattering problem, we show that the transmission across the barrier exhibits {Klein-tunneling behavior protected at normal incidence by the nodal topology}, despite the extended nodal-line dispersion, which can be traced back to Berry-curvature-induced momentum locking. Using the Landauer-Büttiker formalism, we derive general expressions for the linear and nonlinear conductances, including both longitudinal and Hall components, and evaluate them at zero and finite temperature. Our analytical and numerical results elucidate the dependence of the conductance on barrier height and width, as well as on a $\mathcal{PT}$-breaking mass term. We identify distinct transport regimes in which nonlinear contributions are strongly enhanced and transverse Hall currents emerge, providing clear transport signatures of nodal-line topology and suggesting potential routes toward device applications.
\end{abstract}

\maketitle

\section{Introduction} \label{Intro}

The study of topological materials (from insulators to semimetals) has become a central topic in condensed matter physics. These systems give new insights into the classification of electronic states based on their topological robustness and the effects of the underlying Berry phase, which lead to novel and distinctive properties \cite{ZHANG2018580}. In recent years, such properties have gained immense attention, mainly because of the continuous development of smaller-sized electronic devices and the search for alternatives to traditional semiconductor technologies \cite{Gilbert2021}. This highlights the importance of characterizing electronic transport of topological semimetals, because not only will it uncover unique potential properties but also it will provide a way to verify the presence of nontrivial topological states in addition to the observation of band topology by ARPES experiments \cite{RevSMTransport}.

Among these topological materials, topological semimetals such as Dirac semimetals, Weyl semimetals, and nodal-line semimetals stand out as systems where the conduction and valence bands cross in momentum space, and this crossing is protected by symmetries against perturbations \cite{Weng_2016}. The relativistic-like behavior of electrons in these semimetals manifests in distinctive transport properties, including extremely large magnetoresistance \cite{Xiong_2015,Liang_2015}, light effective mass and high electron mobilities \cite{Neupane_2014,Lv_2015}, the chiral anomaly \cite{Nielsen_1983,Xiong_2015,Huang_2015}, and the anomalous Hall effect \cite{Burkov_2014,Nagaosa_2010}. Along with the topological robustness against perturbations of their linear dispersion and spin-momentum locking, these properties make topological semimetals promising candidates for the development of dissipationless transport \cite{Ji2024}, spintronic and magnetoelectric devices \cite{Moore2010}, and topological quantum computation \cite{RevSMTransport}. 

In particular, nodal line semimetals (NLSMs) are 3D materials characterized by band crossings in a one-dimensional line, which may form a closed loop within the Brillouin zone protected by mirror reflection symmetries and $\mathcal{PT}$ symmetry \cite{Fang_2016}. In the low-energy approximation, the Hamiltonian of a nodal line semimetal can be written in terms of the Pauli matrices to represent an orbital (or pseudospin) subspace with a linear dependence on the momentum perpendicular to the nodal ring \cite{PRL119_147402}. Given this structure, and based on the previous results obtained for graphene, where the low-energy model led to the appearance of the quantmum relativistic Klein tunneling \cite{Katsnelson2006}, we considered a sample of the NLSM with a constant electrostatic barrier in it, and analyzed the tunneling through it. It is shown that there is an interference pattern with peaks of perfect transmission at Fabry-Pérot resonances and, similarly to the tunneling in graphene at normal incidence, whenever the pseudospin is conserved at the interfaces of the barrier, the barrier becomes transparent to the particles, hence showcasing the Klein tunneling in NLSM. {We emphasize that this protection applies specifically to the normal-incidence channel, where pseudospin conservation prevents backscattering, while for finite transverse momentum the effective mass contribution suppresses perfect transmission.}

Although the transmission probability cannot be directly measured, its effects manifest in the charge transport properties. Under the Landauer-B\"uttiker formalism, a physical extensive quantity, the conductance, is related explicitly with the transmission function of the material \cite{Datta1995}. The derivation of the Landauer-B\"uttiker formula allows a series expansion of the conductance in terms of the applied field, where the nonlinear terms are related to the derivatives of the transmission function \cite{kawabata_nonlinear_2022}. The applied field produces a longitudinal response on the material, but it also produces an anomalous (or Hall) response proportional to the Berry curvature of the system \cite{XiaoNiuRMP2010}, which for the NLSM results in a topological current that would cancel out to zero due to its symmetry, nevertheless it can be measured through momenta filters \cite{Rui2018} and for that reason it will be considered here. In this work, we compute the linear and quadratic conductances for the longitudinal response considering the effects of a $\mathcal{PT}$ symmetry-breaking mass term and the quadratic conductance for the Hall response, which only persists in the presence of the mass term, both responses are treated at zero and finite temperatures, and an additional finite temperature expression within the Sommerfeld approximation is presented. {We find that these nonlinear interference features are strongly sensitive to thermal broadening, implying that their experimental observation requires sufficiently low temperatures.}

{
It is worth noting that nonlinear Hall responses have recently attracted considerable attention in the context of inversion-broken metals, where they are commonly interpreted within a semiclassical Boltzmann framework in terms of a Berry-curvature dipole \cite{SodemannFu2015}, as demonstrated experimentally in bilayer WTe$_2$ \cite{10.1038/s41586-018-0807-6}. In that scenario, a finite local Berry curvature at the Fermi surface gives rise to a dipolar distribution that produces a second-order Hall response in the presence of time-reversal symmetry. In contrast, the nodal-line system considered here preserves combined $\mathcal{PT}$ symmetry in its ideal form, for which the Berry curvature vanishes identically in the bulk. As we discuss below, the nonlinear response studied in this work emerges only upon introducing a small $\mathcal{PT}$-breaking mass term and is analyzed within a coherent Landauer-B\"uttiker transport framework, thereby addressing a distinct physical regime from the semiclassical Berry-dipole mechanism.
}

The structure of this paper is as follows. In Sec.~\ref{nodal_line_model}, we introduce the nodal-line semimetal model and establish the theoretical framework of the problem, emphasizing the role of $\mathcal{PT}$ symmetry and the setup involving a finite electrostatic barrier. Section~\ref{Klein_tunneling_section} is devoted to the analysis of Klein tunneling in nodal-line semimetals, where we derive the transmission probability across the barrier and highlight the impact of Berry-curvature-induced momentum locking. In Sec.~\ref{logitudinal_conductance_section}, we formulate the longitudinal transport problem within the Landauer approach and derive general expressions for both linear and nonlinear conductance contributions, without specifying material-dependent parameters. Section~\ref{Hall_conductance_section} extends this theoretical analysis to the Hall response, treating linear and nonlinear regimes on equal footing and identifying the conditions under which nonlinear Hall effects arise. The general transport formulas obtained in these sections are subsequently evaluated in Sec.~\ref{charge_transport}, where we compute the conductances explicitly using material parameters representative of an isotropized SrAs$_3$-like nodal-line semimetal and analyze their dependence on chemical potential, barrier characteristics, and temperature. Finally, Sec.~\ref{Conclusions} summarizes our main results and outlines possible directions for future research.

\section{Model and theoretical framework} \label{nodal_line_model}

\subsection{Basic Hamiltonian and band structure}

To model the low-energy physics of a $\mathcal{PT}$-symmetric nodal-line semimetal (NLSM), we adopt a two-band Hamiltonian that captures the essential band-touching features along a closed loop in the Brillouin zone. In its simplest form, the conduction and valence bands cross at $k_{\perp}=k_{0}$ in the plane $k_{z}=0$, giving rise to a nodal ring. Following the approach of Ref.~\cite{PRL119_147402}, we take as our starting point the real, two-component Hamiltonian
\begin{align}
    \hat H_{0}=\hbar v q_{\rho} \sigma _{x} + \hbar v k_{z} \sigma_{y}, \label{Hamiltonian0}
\end{align}
where $q_{\rho}=k_{\perp}-k_{0}$, with $k_{\perp}=\sqrt{k_{x}^{2}+k_{y}^{2}}$ the in-plane momentum measured from the ring of radius $k_{0}$. The Pauli matrices $\sigma_{x,y}$ act in the orbital (pseudospin) subspace and $v$ is an effective Fermi velocity. This $\mathcal{PT}$-symmetric model realizes a gapless nodal ring stabilized by a quantized Berry phase accumulated around it.

For later use, we will consider a symmetry-breaking perturbation that tilts the pseudospin out of the equatorial plane and gaps the ring. The simplest choice is an out-of-plane term $m \sigma_{z}$, so that
\begin{align}
    \hat{H} = \hat{H} _{0} + m  \sigma_{z} . \label{Hamiltonian}
\end{align}
This addition generates finite Berry curvature near the former nodal ring and activates anomalous (including nonlinear Hall-like) transport responses, as discussed in Ref.~\cite{PRB98_155125}. The transport analysis below will make explicit where this perturbation is turned on.

Diagonalizing Eq.~(\ref{Hamiltonian}) yields the spectrum
\begin{align}
    E _{s}(\mathbf{k}) = s \sqrt{ \hbar ^{2} v ^{2} ( q _{\rho} ^{2} + k _{z} ^{2} ) + m^{2} } \equiv s  \mathcal{E} _{\mathbf{k}}, 
\end{align}
with band index $s=\pm 1$ for conduction and valence states, respectively. The normalized eigenvectors can be written as
\begin{align}
    \ket{u _{s}(\mathbf{k})} &= \frac{1}{\sqrt{2 \mathcal{E} _{\mathbf{k}} (\mathcal{E} _{\mathbf{k}} - s m)}} 
    \begin{pmatrix} 
        s \hbar v ( q _{\rho} -i k _{z}) \\[4pt]
        \mathcal{E} _{\mathbf{k}} - s m
    \end{pmatrix}. 
\end{align}
These spinors encode the pseudospin texture of the NLSM in momentum space and provide a convenient starting point to analyze its geometric properties.

\subsection{Topological properties}

Within the two-band description the model Hamiltonian (\ref{Hamiltonian}) can be written as $\hat H = \mathbf{d}(\mathbf{k}) \cdot \boldsymbol{\sigma}$, where $\mathbf{d} (\mathbf{k})=(\hbar v q _{\rho}, \hbar v k _{z},m)$. The geometric response is encoded by the unit vector $\hat{\mathbf{d}} = \mathbf{d}/\mathcal E _{\mathbf{k}}$ and its spherical angles: $\cos \theta = m / \mathcal E _{\mathbf{k}}$ and $\psi = \arg (q _{\rho} - i k _{z})$. In this gauge the Berry connection, defined as $\mathbf{A} _{s} (\mathbf{k}) = i \langle u _{s} (\mathbf{k}) \vert \nabla _{\mathbf{k}} u _{s} (\mathbf{k}) \rangle$, takes the simple form $\mathbf{A} _{s} = \frac{s}{2} (1 - \cos \theta) \nabla _{\mathbf{k}} \psi$ \cite{Berry1984,PRL119_147402}, which, upon expressing $(\theta,\psi)$ in terms of $(q_\rho,k_z)$, yields
\begin{align}
    \mathbf{A} _{s} (\mathbf{k}) = \frac{s}{2} \left( 1 - \frac{m}{\mathcal{E} _{\mathbf{k}}} \right) 
    \frac{ - k _{z} \, \hat{\mathbf{e}} _{q _{\rho}} + q _{\rho} \, \hat{\mathbf{e}} _{k _{z}}  }{q _{\rho} ^{2} + k _{z} ^{2}}, \label{Berry_connection}
\end{align}
which explicitly shows the azimuthal winding of the pseudospin around the nodal ring. The associated Berry curvature is purely azimuthal, 
\begin{align}
    \boldsymbol{\Omega} _{s} (\mathbf{k}) =  s \frac{m \hbar ^{2} v ^{2}}{2 \mathcal{E} _{\mathbf{k}} ^{3}} \, \hat{\mathbf{e}} _{k _{\phi}},  \label{Berry_curvature}
\end{align}
and becomes finite only when $m \neq 0$. Equations \eqref{Berry_connection}-\eqref{Berry_curvature} reproduce the vortex-like pattern of the geometric field localized near the nodal ring \cite{Mikitik1999,Burkov2011,PRL119_147402}.

For $m=0$, the pseudospin lies entirely in the $k_{x}k_{y}$-plane, winding by $2\pi$ around the nodal line. This is reflected in the Berry phase accumulated along a closed loop $\mathcal{C}$ encircling the ring:
\begin{align}
    \gamma _{s} (\mathcal{C}) = \oint _{\mathcal{C}}  \mathbf{A} _{s} (\mathbf{k}) \cdot d \mathbf{k}. \label{Berry_phase}
\end{align}
For the valence band ($s = - 1$), the result is a quantized Berry phase $\gamma _{-} = \pi$, which constitutes the topological invariant protecting the nodal line. This quantization directly leads to the emergence of nearly flat drumhead surface states bounded by the projection of the nodal ring on the surface Brillouin zone \cite{PRL119_147402}. Detailed derivations, including the explicit line-integral and the Stokes-theorem evaluation on a transverse disk, are deferred to Appendix \ref{app:BerryPhase}.

When $m \neq 0$, the pseudospin tilts out of the plane, the Berry phase ceases to be quantized (see  Appendix \ref{app:BerryPhase} for details), and the curvature distribution near the former nodal ring provides the microscopic origin of anomalous linear and nonlinear transport coefficients \cite{PRB98_155125}. In this sense, the model offers a minimal platform to explore the interplay between nodal-line topology, barrier-induced scattering, and transport phenomena.

\section{Klein tunneling in nodal-line semimetals} \label{Klein_tunneling_section}

We consider a gate-defined electrostatic barrier oriented along the transport direction $z$, with translational invariance in the transverse $(x,y)$ plane. The scalar potential couples as $V(z)$ to the two-band Hamiltonian of Eq. \eqref{Hamiltonian}, and we set up a stationary scattering problem at fixed energy $E$ with identical NLSM leads on both sides of the barrier. A rectangular profile is used:
\begin{align}
    V(z) = 
    \begin{cases}
        V _{0}, & |z| \leq L/2, \\[5pt]
        0 , & |z| > L/2,
    \end{cases}
    \label{Barrier_profile}
\end{align}
with width $L$ and height $V _{0}$, infinite along $x$ and $y$.

We address transmission across the electrostatic barrier by formulating a stationary scattering problem at fixed energy $E$ for the nodal-line Hamiltonian with a scalar potential $V(z)$ given by Eq. (\ref{Barrier_profile}). Owing to translational invariance in the transverse plane, the transverse crystal momentum $\mathbf{k} _{\perp} = k _{x} \hat{\mathbf{e}}_{x} + k _{y} \hat{\mathbf{e}}_{y}$ is conserved. Equivalently, the radial offset $q _{\rho} = k _{\perp} - k _{0}$ remains a good quantum number throughout the scattering process. Since $m$ is uniform and $V(z)$ is piecewise constant, each region supports plane-wave eigenspinors at the same energy shift $E - V (z)$, propagating in opposite $z$ directions. The general stationary state is thus a superposition of forward- and backward-moving solutions within the same band character, and continuity at the two interfaces determines the reflection and transmission amplitudes.

In the semi-infinite leads $\vert z \vert > L/2$, the energy of the fermion takes the value $E = s \sqrt{ (\hbar v ) ^{2} (q _{\rho} ^{2} + k _{z} ^{2}) + m ^{2} }$, where $s = \mbox{sgn} (E)$. The plane-wave spinors are $\psi _{s} ^{(\xi)} (\mathbf{r}) = \phi _{s} ^{(\xi)} (z) \, e ^{i ( \mathbf{r} _{\perp} \cdot \mathbf{k} _{\perp} - Et ) } $, being $\phi _{s} ^{(\xi)} (z)$ a two-component column vector of the form
\begin{align}
    \phi _{s} ^{(\xi)} (z) = \frac{1}{\sqrt{2 E ( E - m)}} 
    \begin{pmatrix} 
        s \hbar v ( q _{\rho} -i \xi k _{z}) \\[4pt]
        s (E - m )
    \end{pmatrix} \; e ^{i \xi k _{z} z}  ,
\end{align}
where $k _{z} = \sqrt{ \frac{E ^{2} - m ^{2}}{ (\hbar v ) ^{2} } - q _{\rho} ^{2} }$ is the longitudinal wave number. For a given transverse offset $q _{\rho}$, longitudinal propagation requires a real $k _{z}$, i.e. $E ^{2} - m ^{2} > ( \hbar v q _{\rho} ) ^{2} $; otherwise $k _{z}$ is purely imaginary and the mode is evanescent. Here, $\xi = +$ corresponds to a state carrying positive group velocity along $z$ (for propagating modes), while $\xi = -$ denotes the counterpropagating partner within the same quasiparticle branch $s$. In the following we use $\xi = \pm $ to distinguish  forward- and backward-moving components at fixed energy $E$. 

In the central region $\lvert z \rvert \leq L/2$, quasiparticles experience a uniform electrostatic offset $V$, so the stationary problem at total energy $E$ is posed at the locally shifted energy $E = V_{0} +s' \sqrt{ (\hbar v ) ^{2} (q _{\rho} ^{2} + k _{z} ^{\ast 2}) + m ^{2} } $, where $ { s' = \mbox{sgn} (E-V_{0} -m) } $. The corresponding plane-wave spinors can be written as $\tilde{\psi} _{s'} ^{(\xi)} (\mathbf{r}) = \tilde{\phi} _{s'} ^{(\xi)} (z) \, e ^{i ( \mathbf{r} _{\perp} \cdot \mathbf{k} _{\perp} - Et ) } $, where the $z$-dependent part is the two-component column 
\begin{align}
    \tilde{\phi} _{s'} ^{(\xi)} (z) = \frac{1}{\sqrt{2 (E - V_{0}) ( E - V_{0} - m)}} 
    \begin{pmatrix} 
        s' \hbar v ( q _{\rho} -i \xi k _{z} ^{\ast}) \\[4pt]
        s' (E - V_{0} - m )
    \end{pmatrix} \; e ^{i \xi k _{z} ^{\ast} z} , 
\end{align}
where $k _{z} ^{\ast} = \sqrt{ \frac{ (E - V _{0}) ^{2} - m ^{2}}{ (\hbar v ) ^{2} } - q _{\rho} ^{2} } $ is the longitudinal wave number. As before, propagating solutions require $k _{z} ^{\ast} \in \mathbb{R}$, which implies $(E-V _{0}) ^{2} - m ^{2} > ( \hbar v q _{\rho} ) ^{2} $. 

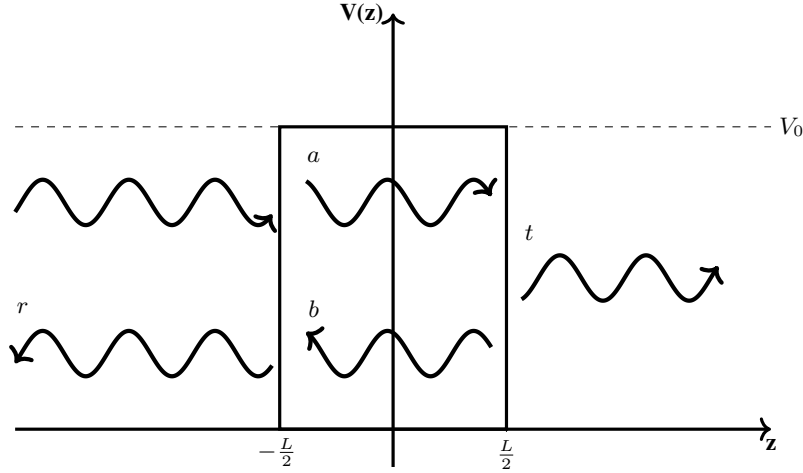
\begin{figure}[h!]
    \centering
     \begin{tikzpicture}
% Definir los ejes
  \draw[very thick,->] (-3.5,0) -- (6.5,0) node[anchor=north] {\textbf{z}};
  \draw[very thick,->] (1.5,-0.5) -- (1.5,5.5) node[anchor=east] {\textbf{V(z)}};
% Marcas sobre los ejes
   \draw (3 cm,1pt) -- (3 cm,-1pt) node[anchor=north] {$\frac{L}{2}$};
   \draw (1pt,0 cm) -- (-1pt,0 cm) node[anchor=north] {$-\frac{L}{2}$};
   \draw[very thin, dashed] (-3.5,4) -- (6.5,4) node[anchor=west] {\textbf{$V_0$}};
% Barrera
   \draw[very thick] (0,0) rectangle (3,4);
% Ondas
   \draw[ultra thick,->,domain=-3.5:-0.1,smooth,samples=100] 
    plot (\x, {0.3*sin(5.51*\x r)+3});
    \draw[ultra thick,<-,domain=-3.5:-0.1,smooth,samples=100] 
    plot (\x, {0.3*sin(5.51*\x r)+1});
    \draw[ultra thick,->,domain=0.35:2.8,smooth,samples=100] 
    plot (\x, {0.3*sin(5.51*\x r)+3});
    \draw[ultra thick,<-,domain=0.35:2.8,smooth,samples=100] 
    plot (\x, {0.3*sin(5.51*\x r)+1});
    \draw[ultra thick,->,domain=3.2:5.8,smooth,samples=100] 
    plot (\x, {0.3*sin(5.51*\x r)+2});
% Etiquetar las regiones y ondas
    \node at (-3.4, 1.6) {$r$};
    \node at (0.45, 3.6) {$a$};
    \node at (0.45, 1.6) {$b$};
    \node at (3.3, 2.6) {$t$};
 \end{tikzpicture}
    \caption{Schematic scattering setup for a nodal-line semimetal (NLSM) in the presence of a gate-defined electrostatic barrier of height $V_{0}$ and width $L$. An incoming state from the left is partially reflected with amplitude $r$ and partially transmitted with amplitude $t$, while the barrier region $z \in [-L/2,\,L/2]$ supports forward- and backward-propagating modes with amplitudes $a$ and $b$. This geometry provides the basis for computing the energy-resolved transmission coefficient $T(E)$ within the Landauer-B\"uttiker framework.}    \label{fig:Tunneling}
\end{figure}

To set up the scattering problem of Fig. \ref{fig:Tunneling}, we expand the stationary solution in each slab as a superposition of the two counterpropagating eigenmodes at the corresponding shifted energy. In the left lead ($z<-L/2$) we choose a unit-flux incoming state and allow for a back-scattered component,
\begin{align}
    \Psi_{\mathrm{L}}(\mathbf{r}) \;=\; \psi_{s}^{(+)}(\mathbf{r}) \;+\; r\,\psi_{s}^{(-)}(\mathbf{r}) ,
\end{align}
with $r$ the reflection amplitude. Inside the barrier ($|z|\le L/2$), the spinor is written as a linear combination of forward- and backward-propagating solutions of the same branch $s'=\mathrm{sgn}(E-V_{0})$,
\begin{align}
    \Psi_{\mathrm{C}}(\mathbf{r}) \;=\; a\,\tilde{\psi}_{s'}^{(+)}(\mathbf{r}) \;+\; b\,\tilde{\psi}_{s'}^{(-)}(\mathbf{r}) ,
\end{align}
where the coefficients $a$ and $b$ will be fixed by the continuity conditions at $z=\pm L/2$. In the right lead ($z> L/2$) only an outgoing wave is present,
\begin{align}
    \Psi_{\mathrm{R}}(\mathbf{r}) \;=\; t\,\psi_{s}^{(+)}(\mathbf{r}) ,
\end{align}
with $t$ the transmission amplitude.

Continuity of the two-component spinor at the interfaces $z = \pm L/2$ imposes the matching conditions
\begin{align}
    \Psi_{\mathrm{L}}(-L/2)=\Psi_{\mathrm{C}}(-L/2),\qquad
    \Psi_{\mathrm{C}}(L/2)=\Psi_{\mathrm{R}}(L/2),
\end{align}
which, at fixed energy $E$ and transverse offset $q _{\rho}$, yield a $4 \times 4$ linear system for the coefficients $\{r,a,b,t\}$. Solving this system provides the transmission amplitude $t (E , \mathbf{k} _{\perp} )$, which yields the transmission probability
\begin{align}
    \mathcal{T} (E , \mathbf{k} _{\perp} ) = \vert t (E , \mathbf{k} _{\perp} ) \vert ^{2} =  \frac{1}{ \cos ^{2} \left( k _{z} ^{\ast} L \right) + \mathcal{R} ^{2} (E,q _{\rho} ) \, \sin ^{2}  \left(k _{z} ^{\ast} L \right) } , \label{transmission_probability}
\end{align}
where
\begin{align}
    \mathcal{R} (E , \mathbf{k} _{\perp} ) = \frac{(\hbar v q _{\rho} ) ^{2} + m ^{2} - E (E - V _{0} )}{(\hbar v) ^{2} k _{z} ^{\ast} k _{z}} .    \label{R_function_def}
\end{align}
To interpret Eq.~\eqref{transmission_probability}, we first delineate the parameter regime in which both longitudinal wavevectors are real, so that transport proceeds via propagating modes rather than evanescent decay. In that regime, the oscillatory factors $\cos(k_z^\ast L)$ and $\sin(k_z^\ast L)$ encode phase accumulation inside the barrier, while the interface mismatch $\mathcal R$ in Eq.~\eqref{R_function_def} quantifies pseudospin/velocity misalignment across the junction. To make explicit the domain where this description applies, we note that longitudinal propagation in the leads and in the barrier requires
\begin{align}
    E^{2}-m^{2} &> (\hbar v q_\rho)^{2},\qquad
    (E-V_{0})^{2}-m^{2} > (\hbar v q_\rho)^{2},
\end{align}
so that $k_{z}, k_{z} ^{\ast} \in \mathbb{R}$. {The transverse momentum $q_\rho$ enters the dispersion in the same quadratic form as the mass term, effectively acting as an additional mass contribution that progressively suppresses perfect transmission away from normal incidence.}
Because the transverse offset $q_\rho$ is conserved, these inequalities define a propagation window in transverse momentum:
\begin{align}
    q _{\rho} ^{2} < q _{\rho,\max} ^{2} \equiv \min \left\{\frac{E ^{2} - m ^{2}}{(\hbar v)^{2}} , \ \frac{(E-V _{0}) ^{2} - m ^{2}}{(\hbar v) ^{2}} \right\} .    \label{window_offset}
\end{align} 

\begin{figure}
    \centering
    \includegraphics[width=0.45\linewidth]{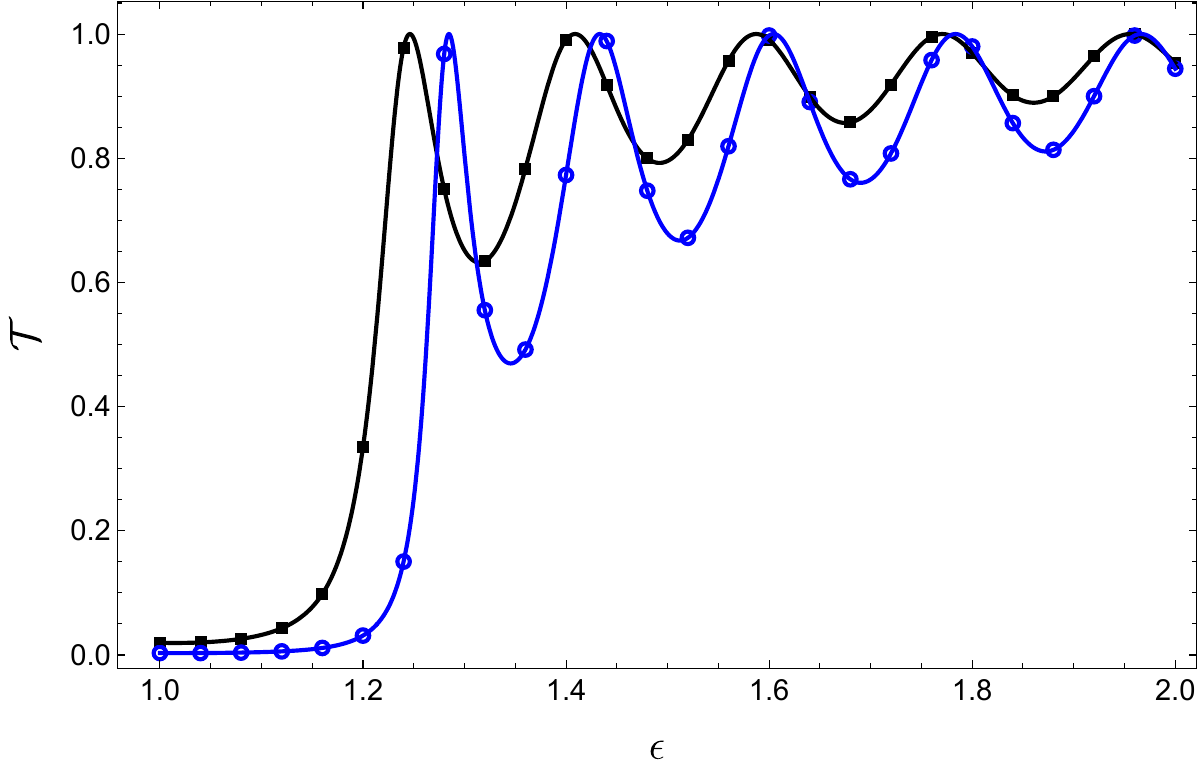}
    \caption{Transmission probability as a function of the energy ratio $\epsilon=\frac{V_0}{E}$ for a fixed value of perpendicular momentum $k_{\perp}=0.96~k_0$ considering an energy of $E=35~\text{meV}$, a barrier width $L=100~\text{nm}$, and the realistic values of $\hbar v \simeq 2.1~\mathrm{eV \cdot \text{\AA}}$, $k_{0} \simeq 0.066~\text{\AA}^{-1}$, which are described in Sec.~\ref{charge_transport}. The black curve corresponds to the gapless nodal-line semimetal, whereas the blue curve include a mass term ($m = 5~\text{meV}$). Both cases exhibit the Fabry-P\'erot resonances, while the opening of the energy gap raises the spectral threshold.}
    \label{KleinTun}
\end{figure}

Geometrically, rotational symmetry around $z$ maps Eq.~\eqref{window_offset} to a cone in momentum space: introducing the incidence angle $\alpha$ defined by $q _{\rho} = k _{\mathrm F} \sin \alpha $ and $k _{z} = k _{\mathrm F} \cos \alpha $, where $k _{\mathrm F} = \sqrt{E ^{2} - m ^{2}}/(\hbar v)$ is the Fermi wave-vector, barrier propagation exists only for
\begin{align}
    \sin ^{2} \alpha  \leq  \frac{(E-V_{0})^{2}-m^{2}}{E^{2}-m^{2}} \equiv  \sin ^{2} \alpha _{c} ,
\end{align}
i.e. $\alpha \leq \alpha _{c}$. As $\vert E - V _{0} \vert $ approaches $m$, the acceptance cone shrinks and closes at $(E - V _{0}) ^{2} = m ^{2}$, beyond which all channels are evanescent in the barrier. Note that the angle $\alpha$ is defined in the terms of the transverse momentum $q _{\rho} = k _{0} - k _{\perp}$, and then it does not corresponds to the physical incidence angle. {In contrast to strictly two-dimensional massless Dirac fermions, the protection of Klein tunneling in this three-dimensional nodal-line system is therefore restricted to the $q_\rho=0$ channel, while oblique channels acquire an effective mass and exhibit reduced transmission. It is worth emphasizing a structural difference with respect to graphene. 
In two-dimensional Dirac systems, the conserved transverse momentum $k_y$ enters the longitudinal scattering problem as an effective mass term, and the special channel $k_y=0$ coincides with the Dirac point itself, where the spectrum is globally gapless. 
In contrast, in a nodal-line semimetal the protected channel corresponds to $q_\rho = k_\perp - k_0 = 0$, namely to momenta lying on the projection of the nodal ring rather than at the center of the Brillouin zone. 
When a small $\mathcal{PT}$-breaking mass $m$ is present, the spectrum remains genuinely gapped even at $q_\rho=0$, so the persistence of the normal-incidence channel does not stem from eliminating a true gap but from the winding of the pseudospin texture along the extended nodal loop. 
This geometric origin of channel-selective Klein tunneling is therefore specific to nodal-line semimetals and has no direct analogue in graphene’s pointlike Dirac structure.}

Physically, the propagation window reflects the tradeoff between transverse kinetic energy, which grows with $q _{\rho}$, and the longitudinal kinetic energy available to sustain motion along $z$. A finite mass $m$ raises the spectral threshold and tightens the window; the electrostatic barrier shifts the local energy from $E$ to $E - V _{0}$, further restricting propagation inside the central region. In the massless limit ($m = 0$), the barrier condition reduces to $ \vert E - V _{0} \vert >\hbar v \vert q _{\rho} \vert$, so at fixed $q _{\rho}$ propagation is lost when the gate nearly cancels the incident energy. 
{This behavior is clearly illustrated in Fig.~\ref{KleinTun}, where a fixed value of $k_{\perp}$ (and hence a fixed $q_{\rho}$) is considered. When the barrier energy $V_0$ is comparable to the incident energy $E$, i.e., $\frac{V_0}{E}\simeq 1$, the transmission probability vanishes. Once the energy ratio exceeds the spectral threshold, the transmission starts and the Fabry-P\'erot resonances begin to appear.}

It is instructive to organize the discussion around several limiting regimes of Eqs.~\eqref{transmission_probability}-\eqref{R_function_def}, which expose the competing roles of pseudospin alignment, phase accumulation inside the barrier, and evanescent decay. The cases collected below provide (i) exact benchmarks (uniform medium and normal incidence), (ii) the onset of propagation at threshold, (iii) the interference pattern associated with Fabry-Pérot resonances, and (iv) the asymptotics in the tunneling and thin-barrier regimes. Together, they delineate the parameter space where Klein tunneling survives (massless, $q _{\rho} = 0$), clarify how a finite mass $m$ and oblique incidence ($q _{\rho} \neq 0$) suppress transmission, and yield compact formulas that will be used to validate and interpret our numerical results below.

In the absence of a barrier ($V_{0}=0$), the structure reduces to a uniform NLSM and the interface matching is trivial. The central region is identical to the leads, so there is no pseudospin or velocity mismatch and the transfer matrix collapses to a pure phase accumulation across the length $L$: $r=0$ and $t = e ^{i k _{z}L}$. Consequently, $R=0$ and the transmission is unity for all channels, $\mathcal{T} = 1$, as expected for a homogeneous medium.

At normal incidence ($q _{\rho} = 0$) the pseudospin of the massless nodal-line quasiparticle lies entirely in the $\sigma _{y}$-axis and is locked to the longitudinal momentum. Because the barrier enters as a scalar potential, it does not couple to pseudospin; consequently, the incident and transmitted spinors remain collinear across each interface, whereas a reflected component would require a pseudospin flip and is therefore forbidden. This chirality constraint is encoded by the interface mismatch parameter $\mathcal{R}$ above: for $m=0$ one finds $\mathcal{R} ^{2} = 1$, which collapses the denominator of Eq.~\eqref{transmission_probability} to unity and yields $\mathcal{T} = 1$ for any $V _{0}$ and $L$ (perfect Klein tunneling). {The robustness of Klein tunneling identified here should thus be understood as the topology-protected stability of the normal-incidence channel ($q_\rho=0$), rather than as angle-independent perfect transmission across all transverse modes.}
Physically, perfect transmission persists regardless of carrier inversion under the barrier (electron-hole conversion when $s \neq s'$). For $m \neq 0$, the spinor acquires a finite $\sigma _{z}$ component, pseudospin is no longer strictly conserved at the interfaces, and backscattering becomes allowed: $\mathcal{T}$ drops below unity except at \textit{Fabry-Pérot resonances} $k _{z} ^{\ast} L = n \pi$. Propagation within the barrier further requires $(E - V _{0}) ^{2} > m ^{2}$; otherwise the internal mode is evanescent and $\mathcal{T}$ is exponentially suppressed.

The near-threshold regime corresponds to the onset of propagation for the longitudinal mode inside the barrier, $k _{z} ^{\ast} \to 0$, i.e., $\vert ( E - V _{0}) \vert \to \sqrt{m^2+(\hbar v q _{\rho}) ^{2} }$. In this limit the Fabry-Pérot oscillations disappear and Eq.~\eqref{transmission_probability} reduces to a Lorentzian in the mismatch parameter $\eta$,
\begin{align}
    \mathcal{T} \xrightarrow[k _{z} ^{\ast} \to 0]{} \frac{1}{1 + \eta ^{2} } , \qquad \eta \equiv \frac{L}{(\hbar v) ^{2} \, k _{z} } \Big[ ( \hbar v q _{\rho} ) ^{2} + m ^{2} - E(E-V _{0} ) \Big] .
\end{align}
The quantity $\eta$ encodes the combined pseudospin and velocity mismatch at the interfaces: it grows linearly with the barrier width $L$ and diverges as the external longitudinal momentum $k_z$ approaches its own cutoff (grazing incidence), thereby enhancing reflection. An entirely analogous expression holds when the threshold is approached on the lead side ($k_z\to 0$). Notably, for massless quasiparticles at normal incidence ($m=0$, $q_\rho=0$) one has $\eta\to 0$ even at threshold $E \to V _{0}$, so $\mathcal{T} \to 1$ and Klein tunneling remains perfect; away from that special line (finite $m$ and/or $q _{\rho} \neq 0$), $\eta$ becomes finite and the transmission is reduced according to the above $1/(1 + \eta ^{2} )$ law.

In the tunneling regime the longitudinal mode inside the barrier is nonpropagating, $(E - V _{0} ) ^{2} - m ^{2}<(\hbar v q _{\rho} ) ^{2}$, and the longitudinal momentum becomes purely imaginary, $k _{z} ^{\ast} = i \kappa$ with $\kappa > 0$. The barrier then acts as a classically forbidden region of decay length $\kappa ^{-1}$: the Fabry-Pérot oscillations are replaced by hyperbolic functions, $\cos(k _{z} ^{\ast} L) = \cosh( \kappa L)$ and $\sin(k _{z} ^{\ast} L) = i \sinh(\kappa L)$, and the interface mismatch enters through $\mathcal R = i \widetilde{\mathcal R}$ with $\widetilde{\mathcal R} = [(\hbar v q _{\rho} ) ^{2} + m ^{2} - E ( E - V _{0} )]/[(\hbar v) ^{2} \, \kappa \, k _{z} ]$. Inserting these relations into Eq.~\eqref{transmission_probability} yields
\begin{align}
    \mathcal{T} =  \frac{1}{\,\cosh^{2}(\kappa L)+\widetilde{\mathcal{R}}^{2}\sinh^{2}(\kappa L)\,}
    \xrightarrow[\kappa L\gg 1]{}\; \frac{4\,e^{-2\kappa L}}{1+\widetilde{\mathcal{R}}^{2}} ,
\end{align}
which exhibits the expected exponential suppression with barrier width $L$ and decay constant $\kappa$, modulated by the pseudospin/velocity mismatch prefactor $(1 + \widetilde{\mathcal R} ^{2}) ^{-1}$.

In the thin-barrier regime the phase accumulated across the central slab is small, $k _{z} ^{\ast} L \ll 1$. Keeping terms up to $\mathcal O \big((k _{z} ^{\ast} L) ^{2}\big)$ in \eqref{transmission_probability} (i.e., expanding both $\sin$ and $\cos$) one finds
\begin{align}
    \mathcal{T} \simeq  \frac{1}{ 1 + \big(\mathcal R ^{2} - 1 \big)\,(k _{z} ^{\ast} L) ^{2} \,}
     \simeq  1 - \big( \mathcal R ^{2} - 1 \big) \, (k _{z} ^{\ast} L) ^{2} + \mathcal{O} \big((k _{z} ^{\ast} L) ^{4} \big) .
\end{align}
This form makes explicit that the departure from perfect transmission is quadratic in $L$ and controlled solely by the interfacial pseudospin/velocity mismatch through $\mathcal{R}$. Importantly, for massless quasiparticles at normal incidence ($m = 0$, $q _{\rho} = 0$) one has $\mathcal R ^{2} = 1$, so the $\mathcal{O}  \big((k _{z} ^{\ast} L) ^{2} \big)$ correction vanishes and $\mathcal{T}$ remains unity to this order, consistent with perfect Klein tunneling. The expression above applies when the mode in the barrier is propagating (real $k _{z} ^{\ast} $); in the evanescent case ($k _{z} ^{\ast} = i \kappa$) the thin-barrier expansion should instead be carried out with $\kappa L \ll 1$.

\section{Landauer formalism for nonlinear charge transport} \label{transport_section}

\subsection{Longitudinal conductance} \label{logitudinal_conductance_section}

The transmission probability, and in particular the Klein-tunneling physics uncovered above, is not measured directly; instead it is inferred from charge transport. Among several formalisms (Boltzmann, Kubo, NEGF), we adopt the Landauer-Büttiker picture, which treats the sample as a phase-coherent scatterer between two large reservoirs and provides a transparent link between energy-resolved transmission and current \cite{Landauer1957,Landauer1970,Buttiker1986,Datta1995,Beenakker1997}. Within this framework the reservoirs are in local equilibrium at temperature $T$ with Fermi functions
\begin{align}
    f _{\mathrm L}(E) = f _{\mathrm{eq}}(E - \mu _{\mathrm L}), \qquad f _{\mathrm R}(E) = f _{\mathrm{eq}}(E - \mu _{\mathrm R}) ,
\end{align}
and the bias $V$ enters through the electrochemical potentials. For generality we parametrize the bias partition as
\begin{align}
    \mu _{\mathrm L} = \mu + \eta \, eV , \qquad    \mu _{\mathrm R} = \mu - (1 - \eta ) \, eV , \label{eq:bias_partition}
\end{align}
with $\eta \in [0,1]$. For context, alternative transport approaches used in the literature include the Kubo formalism and Boltzmann transport theory \cite{Kubo1957,Ziman1960}, as well as nonequilibrium Green's functions \cite{HaugJauho2008,Datta1995}.
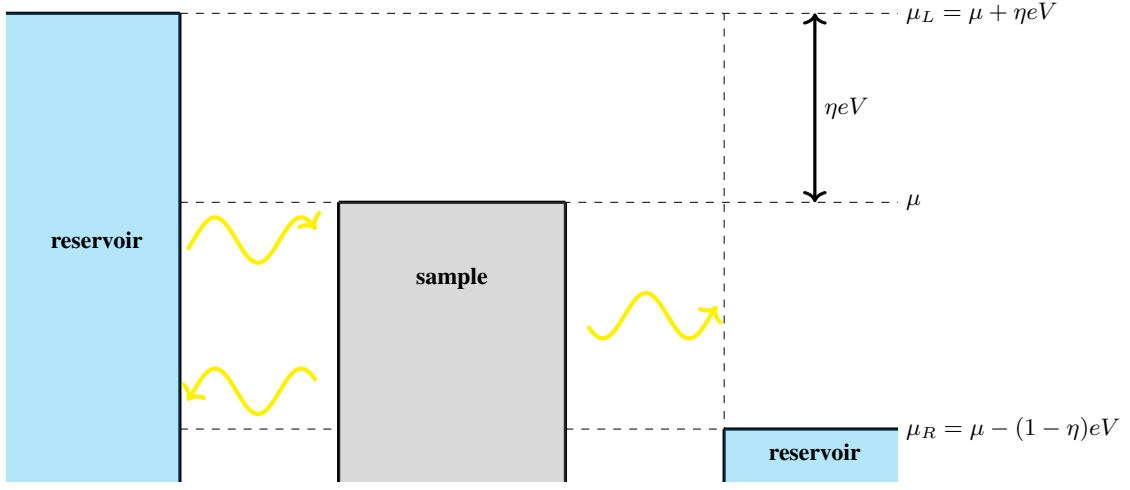
\begin{figure}[h!]
    \centering
    \begin{tikzpicture}
% Definir los ejes
  \draw[very thick,-] (3.6,0) -- (5.9,0);
  \draw[very thick,-] (-3.6,-0.7) -- (-3.6,5.5);
   \draw[very thick,-] (3.6,-0.7) -- (3.6,0);
  
  \draw[very thick,<->] (4.8,3) -- (4.8,5.5)node[midway, right] {\textbf{$\eta eV$}};

   \draw[very thin, dashed] (-3.6,5.5) -- (5.9,5.5)node[anchor=west] {\textbf{$\mu_L=\mu+\eta eV$}};
   \draw[very thin, dashed] (-3.6,0) -- (-1.5,0);
   \draw[very thin, dashed] (1.5,0) -- (5.9,0)node[anchor=west] {\textbf{$\mu_R=\mu-(1-\eta)eV$}};
   \draw[very thin, dashed] (-3.6,3) -- (5.9,3)node[anchor=west] {\textbf{$\mu$}};
   \draw[very thin, dashed] (3.6,-0.5) -- (3.6,5.5);
% Barrera
   \draw[very thick,-] (-1.5,-0.7) -- (-1.5,3);
   \draw[very thick,-] (1.5,-0.7) -- (1.5,3);
   \draw[very thick,-] (-1.5,3) -- (1.5,3);
   \draw[very thick,-] (-5.9,5.5) -- (-3.6,5.5);

   \fill[cyan, opacity=0.3] (-5.9,-0.7)-- (-3.6,-0.7)  -- (-3.6,5.5) -- (-5.9,5.5);
   \fill[cyan, opacity=0.3] (3.6,-0.7)-- (5.9,-0.7)  -- (5.9,0) -- (3.6,0);
   \fill[gray, opacity=0.3] (-1.5,-0.7)-- (1.5,-0.7)  -- (1.5,3) -- (-1.5,3);
% Ondas
   \draw[ultra thick,->,domain=-3.5:-1.8,smooth,samples=100, yellow] 
    plot (\x, {0.3*sin(5.51*\x r)+2.5});
    \draw[ultra thick,<-,domain=-3.5:-1.8,smooth,samples=100, yellow] 
    plot (\x, {0.3*sin(5.51*\x r)+0.5});
    \draw[ultra thick,->,domain=1.8:3.5,smooth,samples=100, yellow] 
    plot (\x, {0.3*sin(5.51*\x r)+1.5});

    \node at (-4.7, 2.5) {\textbf{reservoir}};
    \node at (0, 2) {\textbf{sample}};
    \node at (4.8, -0.3) {\textbf{reservoir}};
\end{tikzpicture}
    \caption{Schematic illustration of two-terminal transport within the Landauer-B\"uttiker picture. The sample is modeled as a phase-coherent scatterer connected to two large reservoirs, each in local equilibrium with electrochemical potentials $\mu_{L} = \mu + \eta eV$ and $\mu_{R} = \mu - (1-\eta)eV$, respectively. The parameter $\eta \in [0,1]$ specifies how the external bias $V$ is partitioned between the contacts. Incident modes from the left reservoir are partially transmitted through the sample and partially reflected back, and the resulting energy-resolved transmission coefficient $\mathcal{T} (E)$ determines the current through Eq.~(\ref{Landauer_current}).}
    \label{fig:Landauer}
\end{figure}

The two-terminal current is then
\begin{align}
    I(V) = \frac{e}{h} \int _{- \infty} ^{\infty}  \mathcal{T} (E) \, \big[ f _{\mathrm L}(E) - f _{\mathrm R}(E) \big]\;  dE ,   \label{Landauer_current}
\end{align}
which is the standard Landauer-Büttiker result \cite{Landauer1957,Landauer1970,Buttiker1986,Beenakker1997,Datta1995}. In the presence of transverse translation invariance, $\mathcal{T} (E)$ implicitly denotes the sum over propagating transverse channels. Equivalently, one may write \cite{BILP1985,Beenakker1997,Datta1995,NazarovBlanter2009,FisherLee1981}
\begin{align}
    \mathcal{T} (E) = \sum_{\mathbf k_{\perp}} \; \mathcal{T} (E , \mathbf{k} _{\perp} )  ,   \label{mode_sum}
\end{align}
where the summation runs over all transverse momenta $\mathbf k _{\perp}$.

{\color{red}
The Landauer-B\"uttiker framework employed here describes coherent ballistic transport through a finite scattering region, where the current is determined by the transmission probability of quantum states injected from equilibrium reservoirs. In this regime, transport is governed by quantum interference and resonant scattering processes, without introducing a phenomenological relaxation time. This contrasts with semiclassical Boltzmann transport theory, which describes diffusive nonequilibrium transport in extended systems and where nonlinear response coefficients explicitly depend on a finite relaxation time $\tau$. Consequently, the nonlinear conductances derived below should be interpreted as coherent scattering coefficients associated with the energy dependence of the transmission function, rather than as bulk semiclassical nonlinear conductivities.

}

Expanding Eq. \eqref{Landauer_current} for weak bias yields a power series $I(V) = \sum _{n \geq 1} G _{n} V ^{n}$ with coefficients
\begin{align}
    G _{n} (\mu , T) = \frac{e ^{n+1}}{h} \frac{1}{n!} \int _{-\infty} ^{\infty} \; \mathcal{T} (E) \; \Xi _{n} ^{(\eta)}(E) \, d E , \qquad
    \Xi _{n} ^{(\eta)} (E) = \big[(- \eta) ^{n} - ( 1 - \eta ) ^{n} \big] \; \frac{ d ^{n} f _{\mathrm{eq}}(E-\mu)}{d E ^{n}} .     \label{Gn_conductance}
\end{align}
A convenient and physically transparent form follows by integrating by parts $(n-1)$ times and using that $\partial _{E} ^{m} f _{\mathrm{eq}} \to 0 $ as $ \vert E \vert \to \infty$,
\begin{align}
    G _{n} (\mu , T) =  \frac{e ^{n+1}}{h} \frac{1}{n!} \big[ (-\eta ) ^{n} - (1 - \eta ) ^{n} \big]  \int _{-\infty} ^{\infty}  \frac{ d ^{n-1} \mathcal{T} (E) }{ d E ^{ n-1}} \;  \left( - \frac{ d f _{\mathrm{eq}}(E-\mu)}{d E } \right)  \, dE . \label{n_order_conductance}
\end{align}
At finite temperature within the noninteracting Landauer expansion and for a bias-independent scatterer, the coefficients \eqref{n_order_conductance} depend on the bias partition $\eta$. With the symmetric choice $\mu_{\mathrm L/R}=\mu\pm eV/2$ one finds $G_{2n}=0$ for all $n\ge1$ (the current is strictly odd in $V$), whereas an asymmetric choice (we use $\eta=1$ below) lifts this algebraic cancellation and even orders $G_{2n}$ generally appear as thermal averages of energy derivatives of the transmission over the kernel $-\partial _{E} f _{\mathrm{eq}}$. This $\eta$-dependence reflects a convention of the bias-independent model; a partition-independent (gauge-invariant) even-order response requires the self-consistent bias profile $\mathcal{T} (E) \; \to \;  \mathcal{T} (E;V)$, in which case $G _{2n}$ are governed by the device's characteristic potentials \cite{ChristenButtiker1996,SanchezButtiker2004}. In this work we retain $\eta = 1$ and a bias-independent transmission so that $G _{2n}$ act as a compact diagnostic of spectral asymmetry around $\mu$ within the thermal window: they are strongly suppressed along the Klein line ($m=0$, $q_\rho=0$), and they grow away from that line-particularly near Fabry-Pérot conditions or at the onset of evanescence-while increasing temperature broadens the sampling and smooths these features.

In this work we focus on the first- and second-order conductances of nodal-line semimetals. Specializing Eq.~\eqref{n_order_conductance} to $n=1$ and $n=2$ yields
\begin{align}
    G _{1} (\mu,T) &= \frac{e ^{2}}{h} \int _{-\infty} ^{\infty}  \mathcal{T} (E) \;  \left( - \frac{ d f _{\mathrm{eq}}(E-\mu)}{d E } \right)  \; dE , \label{linear_longitudinal_conductance} \\[4pt]   G _{2} (\mu,T) &= \frac{e ^{3}}{2h} \,  \int _{- \infty} ^{\infty} \frac{d \mathcal{T} (E)}{dE} \;  \left( - \frac{ d f _{\mathrm{eq}}(E-\mu)}{d E } \right)  \; dE , \label{quadratic_longitudinal_conductance}
\end{align}
which gives the conductances at finite temperature and finite chemical potential.

At finite temperature the derivative of the Fermi function acts as a normalized thermal kernel,
\begin{align}
    - \frac{d f _{\mathrm{eq}}(E - \mu )}{dE} = \frac{1}{4 k _{\mathrm B} T} \, \mathrm{sech} ^{2} \left( \frac{E - \mu}{2 k _{\mathrm B} T} \right), \label{heat_kernel}
\end{align}
which is strongly peaked around the Fermi level $ E = \mu $ with width $\sim k_{\mathrm B}T$. In the strict zero-temperature limit this kernel collapses to a delta function, $-\partial_E f_{\mathrm{eq}}\to \delta(E-\mu)$, and Eqs. (\ref{linear_longitudinal_conductance}) and (\ref{quadratic_longitudinal_conductance}) reduce to
\begin{align}
    G _{1} (\mu , T=0) = \frac{e ^{2}}{h} \, \mathcal{T} (E) \Big| _{E = \mu} , \qquad G _{2}  (\mu , T=0 ) = \frac{e ^{3}}{2h} \, \frac{d\mathcal{T} (E)}{dE} \Bigg| _{E = \mu} . \label{long_conductances_zero_temp}
\end{align}
Thus, at $T = 0$ the linear conductance samples the transmission at the Fermi level, while the quadratic coefficient is controlled by the local spectral asymmetry through $\mathcal{T}'(\mu)$.

Temperature effects are essential in interpreting transport measurements: the derivative of the Fermi function (\ref{heat_kernel}) acts as an energy-resolution kernel of width $\sim k _{\mathrm B}T$, which smears sharp spectral features (e.g., Fabry-Pérot resonances), suppresses quantum interference at elevated $T$, and weights responses by local energy derivatives of the transmission \cite{Datta1995,Beenakker1997,Imry2002}. Moreover, finite temperature underlies thermoelectric phenomena-currents become sensitive to particle-hole asymmetry through $T'(E)$, so quantitative comparison with experiment requires retaining the full thermal kernel or, at low $T$, employing the Sommerfeld expansion used below \cite{SivanImry1986}.

At low but finite temperatures we will use the Sommerfeld expansion to include thermal broadening systematically \cite{Sommerfeld1928,Mahan2000,Ziman1960,AshcroftMermin}. For a smooth energy dependence of the transmission near the Fermi level, the first two conductances become
\begin{align}
    G _{1}(\mu,T) & \simeq \frac{e ^{2}}{h} \left[
        \mathcal{T}(E ) + \frac{\pi ^{2}}{6}\,(k _{\mathrm B}T ) ^{2} \, \mathcal{T}''(E)        + \frac{7 \pi ^{4}}{360} \, (k _{\mathrm B}T) ^{4} \, \mathcal{T} ^{(4)} (E) + \cdots \right] \Bigg| _{E = \mu} , \\[6pt]    G _{2}(\mu,T) & \simeq \frac{e ^{3}}{2h} \left[        \mathcal{T}'(E) + \frac{\pi ^{2}}{6} \,(k _{\mathrm B}T) ^{2} \, \mathcal{T}''' (E)        + \frac{7 \pi ^{4}}{360} \, (k _{\mathrm B}T) ^{4} \, \mathcal{T} ^{(5)} (E) + \cdots \right] \Bigg| _{E = \mu} .
\end{align}
These expressions show explicitly how temperature broadens the sampling around $E=\mu$: $G_{1}$ is governed by $\mathcal{T}$ and its even derivatives, while $G_{2}$ is controlled by the local spectral asymmetry through odd derivatives of $\mathcal{T}$, with thermal corrections set by powers of $(k_{\mathrm B}T)^{2}$.

\subsection{Hall conductance} \label{Hall_conductance_section}

The transverse (Hall-like) response we study originates from the Berry-curvature correction to the semiclassical equations of motion \cite{SundaramNiu1999,XiaoNiuRMP2010,NagaosaAHE2010}. For a Bloch band with dispersion $E _{s}(\mathbf{k})$ and Berry curvature $\boldsymbol{\Omega}_{s} (\mathbf k)$, the wave-packet dynamics in the presence of electromagnetic fields reads
\begin{align}
    \dot{\mathbf r} _{s} &= \frac{1}{\hbar}\,\nabla_{\!\mathbf k} E _{s}(\mathbf{k}) - \dot{\mathbf k} _{s} \times \boldsymbol{\Omega} _{s}(\mathbf k) , \\[4pt]  
    \hbar \,\dot{\mathbf{k} } _{s} &= - e \mathbf{E} - e \dot{\mathbf r} _{s} \times \mathbf{B} . 
\end{align}
Eliminating $\dot{\mathbf{k} } _{s}$ (and for $\,\mathbf{B} = 0$) one obtains the band velocity $\mathbf{v} _{s}$ plus an anomalous contribution $\boldsymbol{\upsilon} _{s}$
\begin{align}
    \dot{\mathbf{r}}_{s} = \frac{1}{\hbar} \nabla _{\mathbf{k}} E _{s} (\mathbf{k} ) - \frac{e}{\hbar} \mathbf{E} \times \boldsymbol{\Omega} _{s}(\mathbf{k}) \equiv  \mathbf{v} _{s} (\mathbf{k}) + \boldsymbol{\upsilon} _{s} (\mathbf{k}) .
\end{align}
The anomalous term is odd under time reversal and is nonzero only when the local Berry curvature is finite. In our NLSM model, $\boldsymbol{\Omega}_{s}$ lies in the plane perpendicular to the transport axis (Sec.~\ref{nodal_line_model}), so an electric field applied along $\hat{\mathbf{z}}$ produces a velocity deflected in the transverse plane, which is the microscopic origin of the Hall-like current we compute below. Importantly, when the mass parameter vanishes ($m=0$) the Berry curvature is locally zero even though the Berry phase around the nodal ring is quantized; consequently $\boldsymbol{\upsilon} _{s}$ disappears pointwise, consistent with the suppression of Hall-like signals along the Klein line.

In the coherent Landauer regime, the internal driving field is not simply $V/L _{z}$: the electrochemical drop inside the sample depends on the channel transparency. For a perfectly transmitting mode ($\mathcal{T} = 1$) the voltage drop occurs entirely at the contacts and no field develops in the bulk, while for an opaque mode ($\mathcal{T} \ll 1$) one recovers the semiclassical field \cite{Datta1995,Imry2002}. We therefore adopt the interpolating Landauer field
\begin{align}
    \mathbf E _{\mathrm{eff}} = \frac{V}{L _{z}}\,\big( 1 - \mathcal{T} \big)\,\hat{\mathbf{z}}, \label{Landauer_field}
\end{align}
and evaluate the anomalous velocity as $\boldsymbol{\upsilon} _{s} (\mathbf{k}) = -(e/\hbar)\,\mathbf E _{\mathrm{eff}} \times \boldsymbol{\Omega} _{s} (\mathbf k)$. This prescription enforces the correct limits: (i) along the Klein line where $\mathcal{T} \to 1$ the internal field vanishes and so does the Hall-like current; (ii) for $\mathcal{T} \ll 1$ one recovers the standard semiclassical drift weighted by transmission; and (iii) the transverse response is maximized at intermediate transparencies where the product $\mathcal{T} ( 1 - \mathcal{T} ) $ peaks, typically near the onset of barrier propagation or between Fabry-P\'erot resonances. Building on this microscopic picture, we now derive Landauer-type expressions for the nonlinear Hall conductances.

Building on this microscopic picture, we now express the anomalous-velocity contribution to the transverse current in terms of the Bloch modes. For transmitted states within the momentum-space volume $[k_x,k_x+dk_x]\times[k_y,k_y+dk_y]\times[k_z,k_z+dk_z]$, the imbalance between left- and right-moving injections yields a Hall current increment
\begin{align}
    \mathrm{d}I^{\mathrm H}_i
    &= e\,\upsilon _{s,i}(\mathbf{k})\,
       \mathcal{T}(\mathbf{k})\,
       \Big[ f_{\mathrm{L}}\big(E_s(\mathbf{k})\big)
            - f_{\mathrm{R}}\big(E_s(\mathbf{k})\big) \Big]\,
       \frac{L_x L_y\,d^3\mathbf{k}}{(2\pi)^3},
\end{align}
where $\upsilon _{s,i}(\mathbf{k}) = [\boldsymbol{\upsilon} _{s}(\mathbf{k})] _{i}$ is the $i$th component of the anomalous velocity defined above and $f_{\mathrm{L/R}}$ denote Fermi functions for the left/right reservoirs. To leading order in the applied bias, the Landauer field is given by Eq. (\ref{Landauer_field}); so that, after expanding the Fermi functions as
\begin{align}
    f_{\mathrm{L}}\big(E_s(\mathbf{k})\big)
    - f_{\mathrm{R}}\big(E_s(\mathbf{k})\big)
    \simeq -\,eV\,
    \frac{\partial f_{\mathrm{eq}}\big(E_s(\mathbf{k})-\mu\big)}{\partial E_s(\mathbf{k})}
    + \mathcal{O}(V^2),
\end{align}
the contribution of each mode to the Hall current scales as $V^2$ and is weighted by the Landauer kernel $\mathcal{T}(1-\mathcal{T})$.

Integrating over all transmitted modes, the Hall current along the transverse direction $i=x,y$ acquires the Landauer-like form
\begin{align}
    I^{\mathrm H}_i(V)
    &= \int dI^{\mathrm H}_i 
     \simeq \frac{e^3 V^2}{\hbar}\,
    \frac{L_x L_y}{L_z}\,
    \epsilon_{ijz}
    \int_{\mathrm{BZ}}
    \frac{d^3\mathbf{k}}{(2\pi)^3}\,
    \mathcal{T}(\mathbf{k})\big[1-\mathcal{T}(\mathbf{k})\big]\,
    \Omega_{s,j}(\mathbf{k})\,
    \left(-\frac{\partial f_{\mathrm{eq}}\big(E_s(\mathbf{k})-\mu\big)}
               {\partial E_s(\mathbf{k})}\right).
    \label{Hall_current}
\end{align}
Comparing with $I^{\mathrm H}_i(V) = \big[ G ^{\mathrm{H}} _{2} (\mu , T) \big] _{i} \, V ^{2} + \mathcal{O}(V^3)$, we identify the leading (quadratic) Hall conductance
\begin{align}
    \big[ G ^{\mathrm{H}} _{2} (\mu , T) \big] _{i}
    =   \frac{e ^{3}}{\hbar} \,
        \frac{L_x L_y}{L _{z} } \,
        \epsilon _{ijz}
        \int _{\mathrm{BZ}}
        \frac{d ^3\mathbf{k}}{(2\pi)^3}\,
        \mathcal{T}(\mathbf{k})\big[1-\mathcal{T}(\mathbf{k})\big]\,
        \Omega_{s,j}(\mathbf{k})\,
        \left(-\frac{\partial f_{\mathrm{eq}}\big(E_s(\mathbf{k})-\mu\big)}
                   {\partial E_s(\mathbf{k})}\right),
    \label{2n_order_Hall_conductance}
\end{align}
which is independent of the bias-partition parameter in the two-terminal setup \cite{ChristenButtiker1996}, in direct analogy with the two-dimensional formulation of Ref.~\cite{kawabata_nonlinear_2022}. At strictly zero temperature one may replace the thermal kernel by a delta function, $-\partial_E f_{\mathrm{eq}}(E-\mu) \to \delta(E-\mu)$, obtaining a compact Fermi-surface expression
\begin{align}
    \big[ G ^{\mathrm{H}} _{2} (\mu , T = 0) \big] _{i}
    =   \frac{e ^{3}}{\hbar} \,
        \frac{L_x L_y}{L _{z} } \,
        \epsilon _{ijz}
        \int _{\mathrm{FS}}
        \frac{d S_{\mathbf{k}}}{(2\pi)^3 \, 
              \big|\nabla_{\mathbf{k}} E_s(\mathbf{k})\big|}\,
        \mathcal{T}(\mathbf{k})\big[1-\mathcal{T}(\mathbf{k})\big]\,
        \Omega_{s,j}(\mathbf{k}) ,
    \label{2n_order_Hall_conductance_zero_temp}
\end{align}
where $\mathrm{d}S_{\mathbf{k}}$ is an element of the constant-energy surface $E_s(\mathbf{k})=\mu$.

For later use it is convenient to rewrite Eq.~\eqref{2n_order_Hall_conductance} in terms of an energy-resolved kernel. Introducing
\begin{align}
    \mathcal{F}_{j}(E)
    = \int_{\mathrm{BZ}}
      \frac{d ^{3}\mathbf{k}}{(2\pi)^{3}}\,
      \mathcal{T}(\mathbf{k})\big[1-\mathcal{T}(\mathbf{k})\big]\,
      \Omega_{s,j}(\mathbf{k})\,
      \delta\!\big(E-E_{s}(\mathbf{k})\big) ,
    \label{eq:Fj_energy_resolved}
\end{align}
we can recast Eq.~\eqref{2n_order_Hall_conductance} as
\begin{align}
    \big[ G ^{\mathrm{H}} _{2} (\mu , T) \big] _{i}
    =   \frac{e ^{3}}{\hbar} \,
        \frac{L_x L_y}{L _{z} } \,
        \epsilon _{ijz}
        \int_{-\infty}^{\infty} d E\,
        \mathcal{F}_{j}(E)\,
        \left(-\frac{\partial f_{\mathrm{eq}}(E-\mu)}{\partial E}\right).
    \label{eq:G2H_energy_integral}
\end{align}
The derivative of the Fermi function, $-\partial_E f_{\mathrm{eq}}(E-\mu)$, is an even, normalized function sharply peaked at $E=\mu$ with width $\sim k_{\mathrm B}T$ \cite{AshcroftMermin,Ziman1960}. When the Berry-curvature–weighted kernel $\mathcal{F}_{j}(E)$ varies smoothly on that scale, the standard Sommerfeld expansion applies \cite{Mahan2000,Sommerfeld1928} and yields an asymptotic series in even powers of $k_{\mathrm B}T$,
\begin{align}
    \big[ G ^{\mathrm{H}} _{2} (\mu , T) \big] _{i}
    \simeq   \frac{e ^{3}}{\hbar} \,
             \frac{L_x L_y}{L _{z} } \,
             \epsilon _{ijz}
             \left[
                 \mathcal{F}_{j}(E)
                 + \frac{\pi^{2}}{6}(k_{\mathrm B}T)^{2}\,
                   \mathcal{F}_{j}''(E)
                 + \frac{7\pi^{4}}{360}(k_{\mathrm B}T)^{4}\,
                   \mathcal{F}_{j}^{(4)}(E)
                 + \cdots
             \right]_{E=\mu} .
    \label{Sommerfeld_exp_2nd_Hall}
\end{align}
Equation~\eqref{Sommerfeld_exp_2nd_Hall} makes explicit that the $T\to 0$ limit is governed solely by the value of the energy-resolved kernel $\mathcal{F}_{j}(E)$ at the Fermi energy, $\mathcal{F}_{j}(\mu)$, while the leading thermal correction probes its local curvature $\mathcal{F}_{j}''(\mu)$. Physically, increasing temperature broadens the energy window around the Fermi level, smoothing sharp spectral structures (e.g., Fabry-P\'erot oscillations or near-threshold onsets) encoded in $\mathcal{F}_{j}(E)$ and thereby reducing the magnitude of $\big[ G ^{\mathrm{H}} _{2} (\mu , T) \big] _{i}$ when $\mathcal{F}_{j}(E)$ varies rapidly or changes sign across that thermal width \cite{Datta1995,Beenakker1997,Imry2002}. The Sommerfeld series remains accurate as long as $\mathcal{F}_{j}(E)$ is smooth on the scale $k _{\mathrm{B}} T$; if resonances or band edges lie within that window, one should revert to the full thermal integral in Eq.~\eqref{eq:G2H_energy_integral}.

\section{Charge transport in nodal-line semimetals} \label{charge_transport}

We consider a nodal-line semimetal patterned into a slab and contacted by two semi-infinite leads along the transport axis $z$. The sample's lateral dimensions are $L_{x}$ and $L_{y}$. The leads are taken to be the same NLSM subject to a large electrostatic offset $V$; in the formal limit $\vert V \vert \to \infty$ the number of propagating modes in the reservoirs becomes arbitrarily large, ensuring ideal mode injection and collection.

The transmission amplitude for a given energy $E$ and transverse wave-vector $\mathbf{k} _{\perp} =(k_x,k_y) $ is obtained from the barrier problem solved in Sec.~\ref{Klein_tunneling_section}. Under the condition discussed above (large electrostatic offset), the transmission probability reads
\begin{align}
     \mathcal{T} (E,\mathbf{k} _{\perp}) = \frac{1}{ 1 +   \frac{q _{\rho} ^{2} + \left( \frac{m}{\hbar v} \right) ^{2} }{ k _{z} ^{2}} \, \sin ^{2} ( k _{z} L _{z} )} , \label{tranmission-.transport}
\end{align}
with longitudinal wave number $k _{z}  = \sqrt{ \frac{E ^{2} - m ^{2}}{ (\hbar v ) ^{2} } - q _{\rho} ^{2} }$. In what follows we compute the total conductances by summing this mode-resolved transmission over the transverse momenta. To that end, we impose boundary conditions in the $(x,y)$ plane (e.g., periodic or hard-wall) and expand in Fourier modes, which discretize $\mathbf{k} _{\perp}$. Physical observables are then obtained in the wide-sample (thermodynamic) limit $L _{x} , L _{y} \to \infty$, where the discrete sum is well approximated by a continuum integral and the specific choice of boundary condition becomes irrelevant. Writing the cross-sectional area $A = L _{x} L _{y}$, we replace
\begin{align}
    \mathcal{T} (E) = \sum_{\mathbf k_{\perp}} \; \mathcal{T} (E , \mathbf{k} _{\perp} )   \quad \longrightarrow \quad  \frac{A}{(2\pi)^2}\int d^{2}\mathbf k_{\perp} \; \mathcal{T} (E , \mathbf{k} _{\perp} ) , \label{total_tranmission_probability}
\end{align}
where the factor $A / (2 \pi ) ^{2}$ arises from the density of states in the reciprocal space. Rotational symmetry about $\hat{\mathbf z}$ will be exploited by labeling transverse channels with the offset $q_{\rho}=k_{\perp}-k_{0}$ and the azimuthal angle $\phi$ in the $(k _{x} , k _{y})$-plane. Throughout, we evaluate transport by summing (integrating) over all transverse momenta $\mathbf{k} _{\perp}$; no additional restriction (acceptance window) is imposed here, since whether a channel is propagating or evanescent is handled self-consistently by the scattering solution.

We now introduce polar coordinates $k _{x} = k _{\perp} \cos \phi$ and $k _{y} = k _{\perp} \sin \phi$ in the total transmission probability (\ref{total_tranmission_probability}) and perform the angular integral. We next simplify the radial part (so as to make the remaining integral more tractable numerically) by the change of variables $\kappa = k _{\perp} L _{z}$ and introduce the following dimensionless parameters $\epsilon = \frac{E L_{z}}{\hbar v}$, $\Delta = \frac{mL_{z}}{\hbar v}$ and $\kappa _{0} = k _{0} L _{z}$. Therefore we obtain
\begin{align}
    \mathcal{T} (\epsilon ) =  \frac{A}{ 2 \pi L _{z} ^{2} } \int _{0} ^{ \infty }    \frac{\kappa \, d \kappa }{ 1 +   \frac{  ( \kappa - \kappa _{0}) ^{2} + \Delta ^{2} }{   \epsilon ^{2} - \Delta ^{2} - ( \kappa - \kappa _{0}) ^{2}    } \, \sin ^{2} ( \sqrt{ \epsilon ^{2} - \Delta ^{2} - ( \kappa - \kappa _{0}) ^{2}    } )}  . \label{total_tranmission_probability2}
\end{align}
These rescalings reduce parameter stiffness and improve the conditioning of the numerical quadrature. For $\epsilon^{2}-\Delta^{2}-(\kappa-\kappa_{0})^{2}<0$ we continue $k_{z}L_{z}\to i\chi$ with $\chi=\sqrt{(\kappa-\kappa_{0})^{2}+\Delta^{2}-\epsilon^{2}}$, so that $\sin^{2}(i \chi ) =  -\,\sinh^{2}(\chi)$ and the integrand remains real and positive.

For the numerical analysis, we employ the continuum model defined in Eq.~(\ref{Hamiltonian}), using material parameters representative of an isotropized SrAs$_{3}$-like nodal-line semimetal. SrAs$_{3}$ crystallizes in a monoclinic structure (space group $C2/c$), where the low-energy electronic states arise mainly from hybridized As-$p$ and Sr-$d$ orbitals forming extended As$_6$ octahedra. The lattice constant $a \simeq 5.8~\text{\AA}$ provides the characteristic structural scale. The band inversion between these orbitals produces a closed nodal loop centered around the $Y$ point of the Brillouin zone, protected by the combined $\mathcal{PT}$ symmetry and mirror reflection with respect to the $k_{x}$-$k_{y}$ plane. 

First-principles calculations and ARPES measurements show that, in the absence of spin-orbit coupling, SrAs$_{3}$ hosts a single nearly ideal nodal ring close to the Fermi level, with a small spin-orbit-induced gap of only a few meV. In our simulations we take $m = 5~\text{meV}$, a Fermi velocity $v \simeq 3.2\times10^{5}~\mathrm{m/s}$, and a nodal radius $k_{0} \simeq 0.066~\text{\AA}^{-1}$, consistent with experimental observations~\cite{Lv2018_SrAs3}. These parameters yield an energy scale $\hbar v \simeq 2.1~\mathrm{eV \cdot \text{\AA}}$, ensuring that the transport calculations capture the relevant low-energy physics of the nodal ring. In a tight-binding picture, they correspond to comparable nearest-neighbor hopping amplitudes between As-$p_x$ and As-$p_z$ orbitals along the in-plane and out-of-plane directions, which justifies the use of an isotropic effective model for quantitative estimates of conductance and topological response~\cite{Yang2018_NL_TB,Hirayama2017_NLmaterials}.

We consider a finite cubic sample of dimensions $L_x \times L_y \times L_z$, with $L_i \simeq 200~\text{nm}$ on each side, corresponding to approximately $N_i \simeq 345$ lattice sites along each Cartesian direction. Using the parameters above, we obtain the dimensionless quantities $\Delta \simeq 4.83$ and $\kappa_{0} \simeq 132$. The large value $\kappa_{0} \gg 1$ indicates that many Fermi wavelengths fit within the sample thickness, so the motion along $z$ can be treated as quasi-continuous and the spectrum effectively three-dimensional; quantization effects are therefore negligible compared with thermal or disorder broadening. Meanwhile, $\Delta \simeq 4.83$ shows that the small $\mathcal{PT}$-breaking mass opens only a weak gap, preserving the essential nodal-line topology while enabling controlled symmetry breaking in the transport regime.

\begin{figure}
    \centering
    \includegraphics[width=0.45\linewidth]{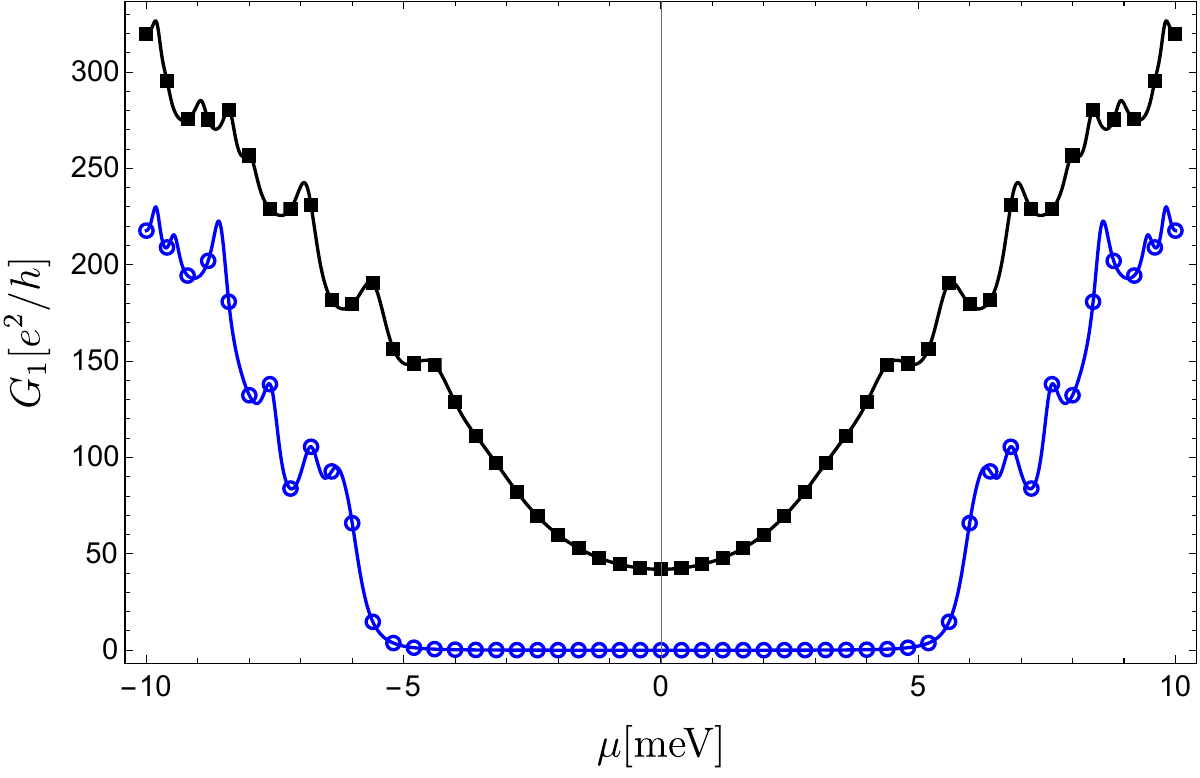} \qquad \includegraphics[width=0.45\linewidth]{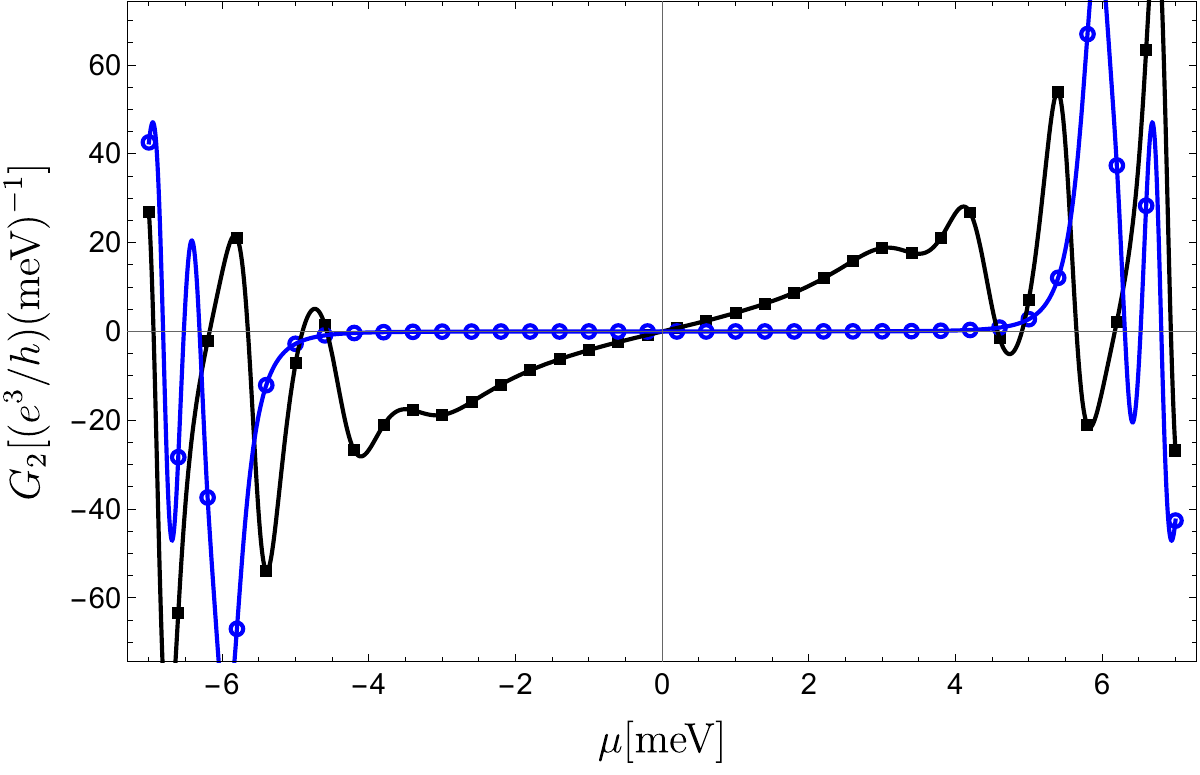}
    \caption{(Color online) Linear ($G_{1}$, left) and quadratic ($G_{2}$, right) conductance densities as a function of the chemical potential $\mu$ at zero temperature. The linear conductance is given in units of $e^{2}/h$, while the quadratic conductance is expressed in units of $e^{3}/h\,(\text{meV})^{-1}$. Black symbols correspond to the gapless nodal-line semimetal, whereas blue symbols include a finite $\mathcal{PT}$-breaking mass term ($m = 5~\text{meV}$). The opening of the gap suppresses the linear conductance and produces a plateau around $\mu \simeq 0$, while the nonlinear response exhibits enhanced oscillations near the band edges. Both $G_{1}$ and $G_{2}$ vanish within the insulating gap, confirming the semimetal-insulator transition.}
    \label{longitudinal_conductances}
\end{figure}

The left panel of Fig.~\ref{longitudinal_conductances} shows the linear conductance density $G_{1}$ (in units of $e^{2}/h$) at zero temperature as a function of the chemical potential $\mu$ (in meV), as defined by Eq. (\ref{long_conductances_zero_temp}). The black curve corresponds to the gapless nodal-line semimetal, while the blue one includes a finite mass term that breaks the $\mathcal{PT}$ symmetry. In both cases, $G_{1}$ increases monotonically with $|\mu|$, reflecting the growth in the number of available propagating modes contributing to the current, as expected from the Landauer formalism~\cite{Datta1995}. The presence of the mass term reduces the conductance values and introduces a plateau-like region around $\mu \simeq 0$, consistent with the opening of an insulating gap where transport is suppressed.

The right panel displays the quadratic conductance density $G_{2}$ (in units of $e^{3}/h\,(\text{meV})^{-1}$), corresponding to the nonlinear (second-order) response. In the gapless regime, $G_{2}$ oscillates around zero and {stabilizes} for large $|\mu|$, indicating that the transmission probability approaches a linear dependence on energy at high chemical potential. In contrast, when a finite mass is introduced, pronounced oscillations emerge near the band edges, where the energy dependence of the transmission changes most rapidly. Both conductance densities vanish when the chemical potential lies within the gap $|\mu| < |m|$, confirming the transition from the semimetallic to the insulating regime. The behavior of $G_{1}$ and $G_{2}$ in the gapless case closely resembles that reported for graphene~\cite{kawabata_nonlinear_2022}, highlighting that similar transport signatures can arise in the three-dimensional nodal-line semimetal despite its distinct topology. {It is worth emphasizing that, although the qualitative behavior of $G_1$ and $G_2$ shares similarities with graphene, particularly the oscillatory structure associated with Fabry-P\'erot interference across a finite barrier, the underlying physical origin in a nodal-line semimetal is fundamentally distinct. In graphene, transport is governed by two discrete Dirac cones located at inequivalent valleys in a two-dimensional Brillouin zone. In contrast, the present system hosts a continuous one-dimensional manifold of band crossings forming a closed nodal loop in three dimensions. The total conductance therefore results from the coherent integration over a continuum of Dirac-like slices parametrized by the azimuthal angle along the ring. This extended phase space of conducting channels, characterized by a finite nodal radius $k_0$, modifies the transverse-mode summation in the Landauer formula and constitutes a genuine three-dimensional feature absent in graphene.}

\begin{figure}
    \centering
    \includegraphics[width=0.45\linewidth]{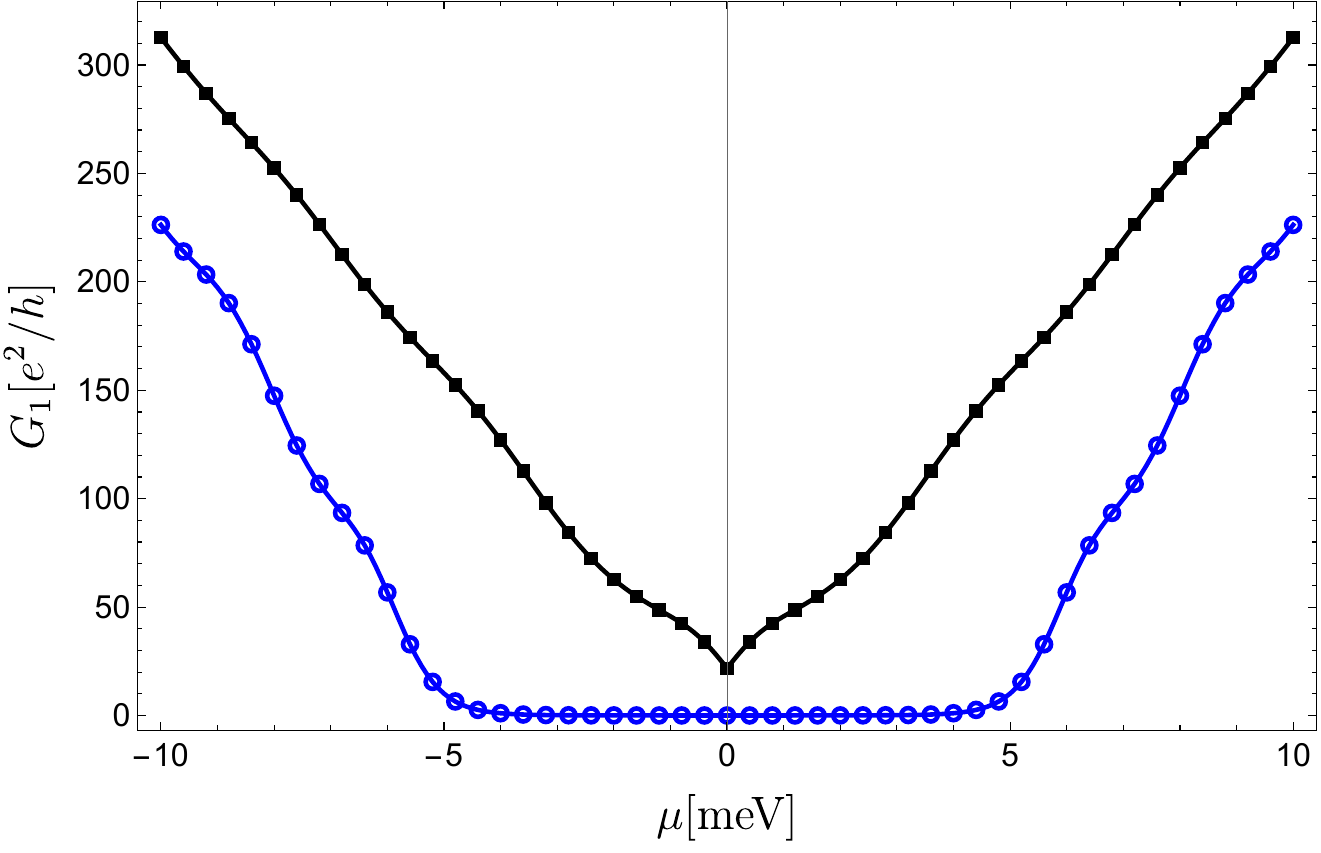} \qquad \includegraphics[width=0.45\linewidth]{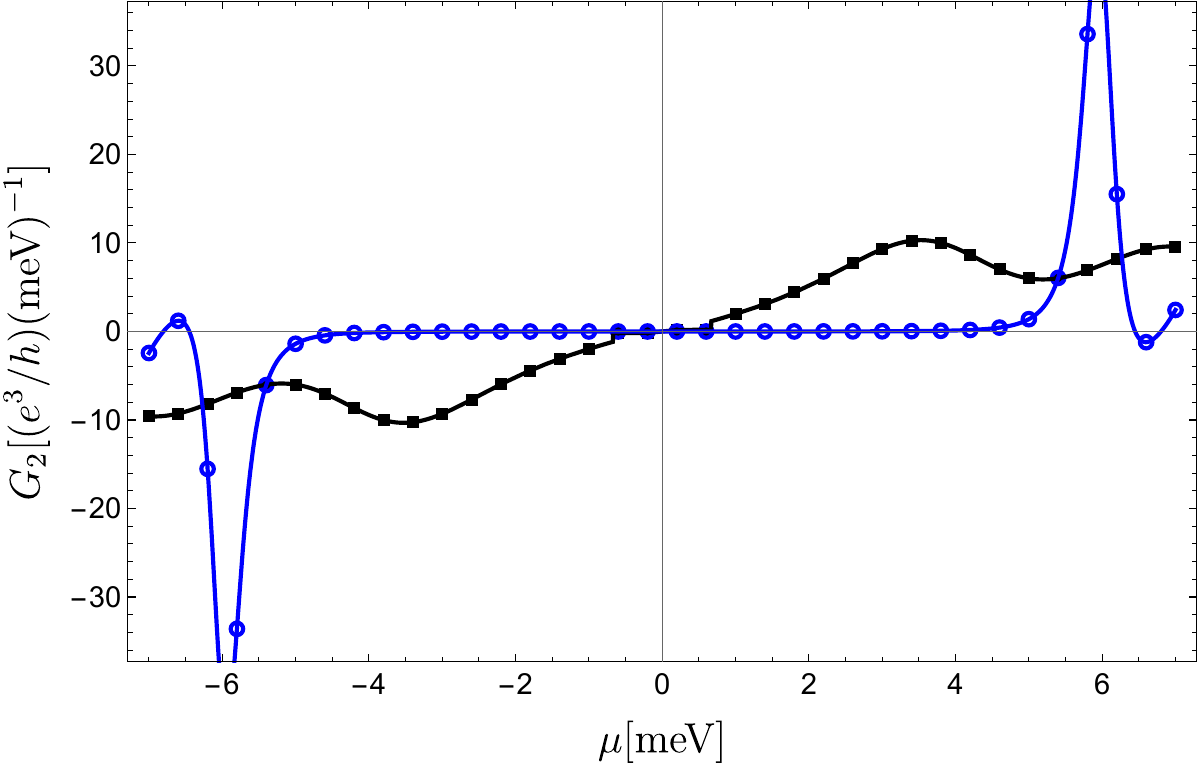}
    \caption{(Color online) Linear ($G_{1}$, left) and quadratic ($G_{2}$, right) conductance densities as a function of the chemical potential $\mu$ at finite temperature (4K). The linear conductance is given in units of $e^{2}/h$, while the quadratic conductance is expressed in units of $e^{3}/h\,(\text{meV})^{-1}$. Finite temperature suppresses the oscillatory behavior of both conductance densities due to the thermal averaging over an energy window of order $k_BT$.
    }
    \label{longitudinal_conductances_FT}
\end{figure}

The finite-temperature conductance densities are shown in Fig.~\ref{longitudinal_conductances_FT}, where the linear (left) and quadratic (right) longitudinal conductances are evaluated at $T=4$ K. In contrast to the zero-temperature case, the oscillatory features observed in both $G_{1}$ and $G_{2}$ are strongly suppressed. This behavior originates from thermal broadening of the electronic distribution, which effectively averages the transmission probability over an energy window of order $k_{B}T$. As a result, sharp energy-dependent structures associated with resonant transmission and mode quantization are smeared out, leading to a smoother dependence on the chemical potential.

For the linear conductance, finite temperature preserves the overall monotonic increase with $|\mu|$, while softening the plateau edges near the gap and reducing the contrast between the gapless and gapped regimes. In the nonlinear response, thermal effects are even more pronounced: the oscillations present at zero temperature are washed out almost completely, and the quadratic conductance exhibits a smooth crossover near the band edges rather than sharp sign-changing features. This reflects the strong sensitivity of nonlinear transport to rapid variations of the transmission with energy, which are particularly vulnerable to thermal averaging. Overall, these results demonstrate that finite temperature acts as an efficient dephasing mechanism for interference-induced oscillations, while leaving the qualitative distinction between gapless and gapped phases intact.

We now turn to the calculation of the Hall conductance given by Eq. (\ref{2n_order_Hall_conductance}) at $T=0$, which involves the transverse response associated with the Berry curvature of the system. To evaluate the corresponding integral, we introduce the change of variables $\boldsymbol{\kappa} = L _{z} \mathbf{k}$ and use the dimensionless variables $\epsilon_{\mu}=\mu L_{z}/\hbar v$, $\Delta=mL_{z}/\hbar v$, and $\kappa_{0}=k_{0}L_{z}$ defined previously. Therefore, the integral we have to evaluate is
\begin{align}
    \big[ G ^{\mathrm{H}} _{2} (\mu , T = 0) \big] _{i}
    =   \frac{e ^{3}}{m h} \,
        \frac{L_x L_y}{L _{z} ^{2} } \,
        \epsilon _{ijz} \,
         \frac{\Delta ^{2}}{2 \epsilon_{\mu} ^{3} } \, 
        \int _{\mathrm{BZ}}
        \frac{d ^{3} \boldsymbol{\kappa} }{(2\pi)^2}\,
        \frac{  \frac{ ( \kappa _{\perp}- \kappa _{0}) ^{2} + \Delta ^{2} }{ \kappa _{z} ^{2}} \, \sin ^{2} ( \kappa _{z}  ) }{ \left[  1 +   \frac{ (\kappa _{\perp}- \kappa _{0}) ^{2} + \Delta ^{2} }{ \kappa _{z} ^{2}} \, \sin ^{2} ( \kappa _{z} ) \right] ^{2}}  \, [ \hat{\mathbf{e}} _{k _{\phi}} ] _{j}  \,  \delta\left( \epsilon_{\mu} - \sqrt{   (\kappa _{\perp} - \kappa _{0}) ^{2} + \kappa _{z} ^{2}  + \Delta ^{2} } \right) . \label{Integral_Hall_conductance_T0}
\end{align}
At this stage, it is convenient to introduce a toroidal coordinate system adapted to the nodal-line geometry of the semimetal. In this parametrization, the momentum components are written as
\begin{align}
    \kappa_x &= (\kappa_0 + \kappa \cos\theta)\cos\phi , \notag \\ \kappa_y &= (\kappa_0 + \kappa \cos\theta)\sin\phi , \notag \\ \kappa_z &= \kappa \sin\theta ,
\end{align}
where $\kappa_0$ denotes the radius of the nodal ring in the $(\kappa_x,\kappa_y)$ plane. The radial coordinate $\kappa$ measures the distance from the nodal line, $\theta$ parametrizes the transverse direction normal to the ring, and $\phi$ is the azimuthal angle along the nodal loop. With this choice, the energy dispersion entering the Dirac delta function depends only on the radial variable through $\kappa_z^{2}+(\kappa_{\perp}-\kappa_0)^2=\kappa^2$, while the momentum-space volume element acquires the Jacobian $J(\kappa,\theta)=\kappa(\kappa_0+\kappa\cos\theta)$. This coordinate system thus provides a natural framework to evaluate the Hall conductance by explicitly resolving fluctuations around the nodal ring. In this coordinate system, equation (\ref{Integral_Hall_conductance_T0}) becomes
\begin{align}
  \hspace{-0.7cm}   \big[ G ^{\mathrm{H}} _{2} (\mu , T = 0) \big] _{i}
    &= \frac{e ^{3}}{m h} \,
        \frac{L_x L_y}{L _{z} ^{2} } \,
        \epsilon _{ijz} \,
         \frac{\Delta ^{2}}{4 \pi \epsilon_{\mu} ^{3} } \, \int _{0} ^{2 \pi } \frac{d \phi}{2 \pi } \, [ \hat{\mathbf{e}} _{k _{\phi}} ] _{j}  \,  
       \int _{0} ^{\kappa _{0}}  \int _{0} ^{2 \pi } 
       d \kappa \, d \theta  \, J (\kappa , \theta ) 
        \frac{  \frac{ \kappa ^{2} \cos ^{2} \theta + \Delta ^{2} }{ \kappa ^{2} \sin ^{2} \theta } \, \sin ^{2} ( \kappa _{z}  ) }{ \left[  1 +   \frac{ \kappa ^{2} \cos ^{2} \theta + \Delta ^{2} }{ \kappa ^{2} \sin ^{2} \theta } \, \sin ^{2} ( \kappa _{z} ) \right] ^{2}}    \,  \delta\left( \epsilon_{\mu} - \sqrt{   \kappa ^{2}  + \Delta ^{2} } \right) .
\end{align}
The $\kappa$-integral can be performed in a simple fashion by using the properties of the Dirac delta. In particular, we use the composition
\begin{align}
    \delta\left( \epsilon_{\mu} - \sqrt{   \kappa ^{2}  + \Delta ^{2} } \right) = \frac{\epsilon_{\mu}}{\sqrt{\epsilon_{\mu} ^{2} - \Delta ^{2}}} \delta  \left( \kappa - \sqrt{\epsilon_{\mu} ^{2} - \Delta ^{2}} \right) .  
\end{align}
After the radial integration we obtain
\begin{align}
\hspace{-1cm}    \big[ G ^{\mathrm{H}} _{2} (\mu , T = 0) \big] _{i}
    &= \frac{e ^{3}}{m h} \,
        \frac{L_x L_y}{L _{z} ^{2} } \,
        \epsilon _{ijz} \,
         \frac{ \Delta ^{2} \kappa _{0} }{ 4 \pi \epsilon_{\mu} ^{2} } \, \Theta (\kappa _{0} - \sqrt{\epsilon_{\mu} ^{2} - \Delta ^{2}} ) \, \int _{0} ^{2 \pi } \frac{d \phi}{2 \pi } \, [ \hat{\mathbf{e}} _{k _{\phi}} ] _{j}  \,  
       \int _{0} ^{2 \pi }  d \theta  \,  
        \frac{  \left( \frac{\epsilon_{\mu} ^{2}}{\epsilon_{\mu} ^{2} - \Delta ^{2}} \csc ^{2} \theta - 1 \right) \, \sin ^{2} ( \sqrt{\epsilon_{\mu} ^{2} - \Delta ^{2}}  \sin \theta  ) }{ \left[  1 +  \left( \frac{\epsilon_{\mu} ^{2}}{\epsilon_{\mu} ^{2} - \Delta ^{2}} \csc ^{2} \theta - 1 \right) \, \sin ^{2} ( \sqrt{\epsilon_{\mu} ^{2} - \Delta ^{2}}  \sin \theta ) \right] ^{2}}   ,  \label{Integral_Hall_conductance_T0_2}
\end{align}
The integrand in Eq.~(\ref{Integral_Hall_conductance_T0_2}) is axially symmetric, except for the tangential component of the Berry curvature, whose angular dependence causes the total angular integral to vanish. Physically, this cancellation arises because every point on the nodal ring has a partner related by inversion symmetry on the opposite side of the loop, leading to opposite contributions to the Hall response. When summing over all azimuthal angles $\phi$, these contributions cancel pairwise, yielding a net zero Hall conductivity, a mechanism analogous to that occurring in graphene, where the two Dirac cones at inequivalent valleys contribute oppositely to the Hall response~\cite{PRB98_155125}.

Guided by this observation, it is instructive to decompose the nodal ring into a continuous family of two-dimensional slices labeled by the azimuthal angle $\phi$. Each slice behaves as an effective graphene-like subsystem with its own local Berry curvature and transverse conductance. From Eq.~(\ref{Integral_Hall_conductance_T0_2}), we therefore define the $\phi$-resolved Hall conductance by performing the integral over the {angular coordinate $\theta$} for fixed $\phi$. Assuming that the chemical potential $E_{F}=\mu$ lies within the conduction band, just above the gap opened by $m\sigma _{z}$, we obtain the angle-dependent Hall conductance density
\begin{align}
    \big[ G ^{\mathrm{H}} _{2} (\mu , T = 0) \big]  ^{\phi}
    &= \frac{e ^{3}}{m h} \,
        \frac{L_x L_y}{L _{z} ^{2} } \,
         \frac{  \Delta ^{2} \kappa _{0} }{4 \pi \epsilon_{\mu} ^{2} } \, \Theta (\kappa _{0} - \sqrt{\epsilon_{\mu} ^{2} - \Delta ^{2}} ) 
       \int _{0} ^{2 \pi }  d \theta  \,  
        \frac{  \left( \frac{\epsilon_{\mu} ^{2}}{\epsilon_{\mu} ^{2} - \Delta ^{2}} \csc ^{2} \theta - 1 \right) \, \sin ^{2} ( \sqrt{\epsilon_{\mu} ^{2} - \Delta ^{2}}  \sin \theta  ) }{ \left[  1 +  \left( \frac{\epsilon_{\mu} ^{2}}{\epsilon_{\mu} ^{2} - \Delta ^{2}} \csc ^{2} \theta - 1 \right) \, \sin ^{2} ( \sqrt{\epsilon_{\mu} ^{2} - \Delta ^{2}}  \sin \theta ) \right] ^{2}}   ,  
\end{align}
such that $ \big[ G ^{\mathrm{H}} _{2} (\mu , T=0) \big] _{i} = \int _{0} ^{2 \pi}  \frac{d \phi}{2 \pi} \, \epsilon _{ijz} \,  \big[ G ^{\mathrm{H}} _{2} (\mu , T=0) \big] ^{\phi} \, [ \hat{\mathbf{e}} _{k _{\phi}} ] _{j}  $.

\begin{figure}
    \centering
    \includegraphics[width=0.45\linewidth]{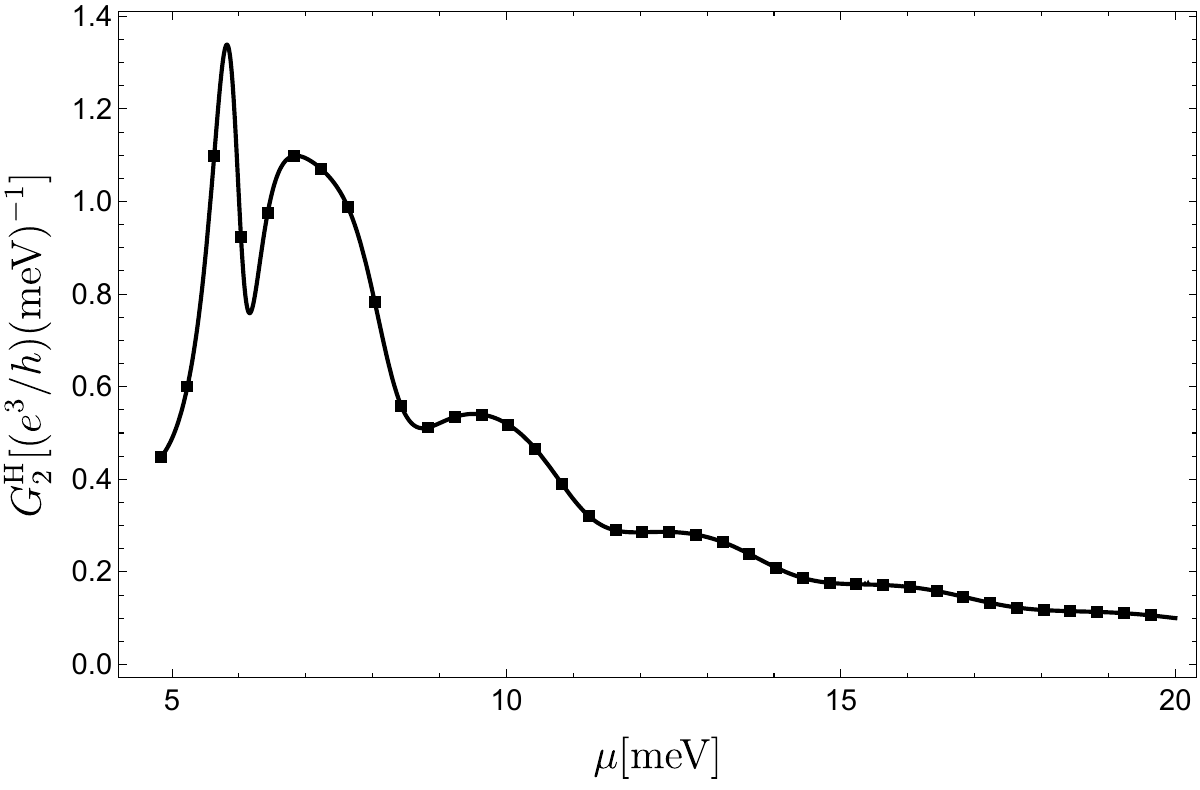}
    \caption{Nonlinear Hall conductance density $[G_{2}^{\mathrm{H}}(\mu , T=0) ] ^{\phi}$ at zero temperature, expressed in units of $e^{3}/h\,(\text{meV})^{-1}$. The signal is predominantly {positive} and displays damped oscillations as $|\mu|$ increases, with extrema located near the band edges where the transmission varies most rapidly. The overall magnitude remains of order unity on the chosen scale, indicating that the nonlinear Hall response is small compared with the linear conductance but clearly discernible.}
    \label{Hall_conductance}
\end{figure}

For the numerical evaluation of the Hall conductance we employ the same material parameters as in the longitudinal case, corresponding to the isotropized SrAs$_3$-like nodal-line semimetal. Using $L_{z} = 200~\text{nm}$ and the previously adopted values of $m$, $v$, and $k_{0}$, the resulting dimensionless parameters are $\Delta \simeq 4.8$ and $\kappa_{0} \simeq 132$. These values ensure that the calculations remain in the weakly gapped regime, where the system preserves its nodal-line character while allowing a finite transverse response to emerge.

Figure~\ref{Hall_conductance} shows the nonlinear Hall conductance density $ G _{2}^{\mathrm{H}}(\mu) $ at zero temperature (units $e^{3}/h\,(\text{meV})^{-1}$). The signal is predominantly positive over the displayed range and exhibits damped oscillations as $|\mu|$ increases, with extrema located near the band edges where the transmission changes most rapidly. The overall magnitude stays of order unity in the chosen units, underscoring that the nonlinear Hall response is much smaller than the linear conductance yet clearly resolvable on the plotted scale. This behavior is consistent with the $\phi$-resolved picture: the Berry-curvature-weighted factor $\mathcal{T}( \mathcal{T} - 1 )$ enhances the response close to onsets of new propagating modes and then decays as the transmission approaches a linear-in-energy regime at large $|\mu|$.

\begin{figure}
    \centering
    \includegraphics[width=0.45\linewidth]{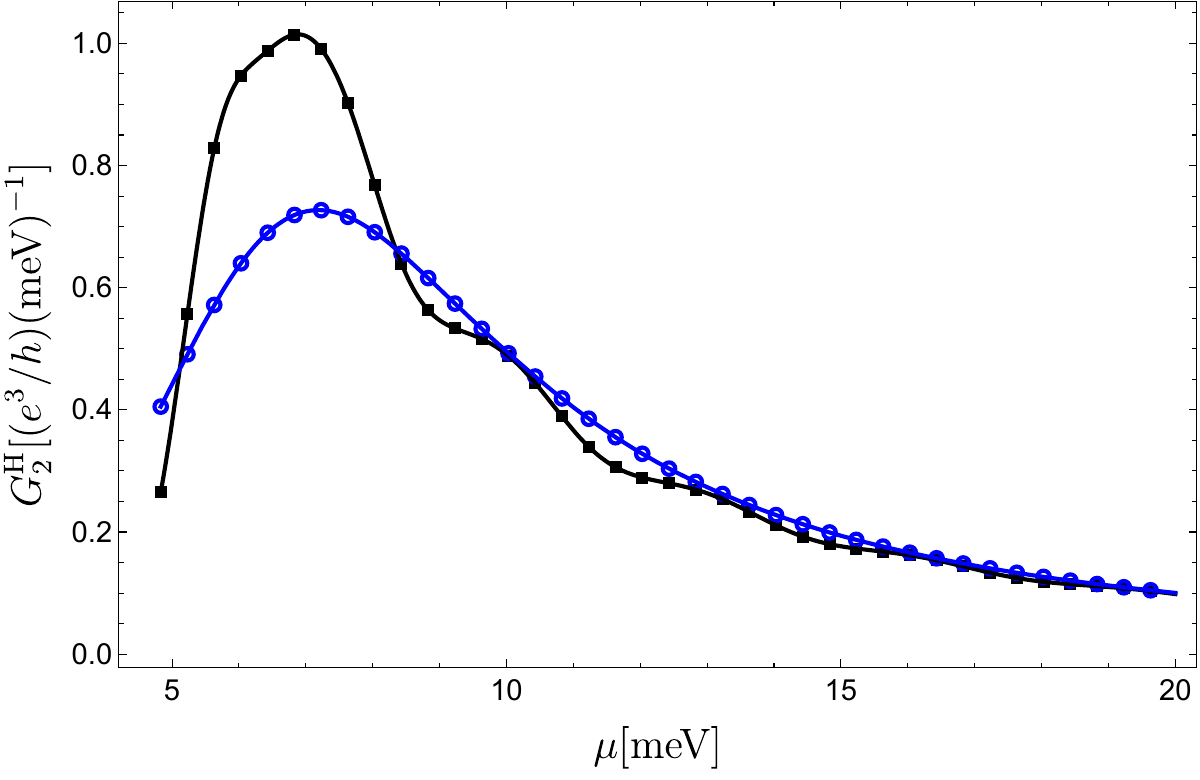}
    \caption{Finite-temperature nonlinear Hall conductance density $[G_{2}^{\mathrm{H}}(\mu , T) ] ^{\phi}$ as a function of the chemical potential $\mu$. The black curve corresponds to $T=3\,\mathrm{K}$, while the blue curve shows the result at $T=12\,\mathrm{K}$. Increasing temperature progressively suppresses the oscillatory features observed at zero temperature, leading to a smoother response and a reduced peak amplitude. The overall positive sign and the monotonic decay at large $\mu$ are preserved, indicating that thermal broadening primarily washes out interference-induced contributions to the nonlinear Hall response.}
    \label{Hall_conductance_finiteT}
\end{figure}

We now turn to the finite-temperature nonlinear Hall conductance, obtained by replacing the zero-temperature energy constraint with the thermally broadened Fermi distribution. In this case, the angle-resolved Hall conductance reads
\begin{align}
    \big[ G ^{\mathrm{H}} _{2} (\mu , T) \big] ^{\phi}
    =   \frac{e ^{3}}{h} \,
        \frac{L_x L_y}{L _{z} ^{2} } \, \frac{ \Delta ^{2} \kappa _{0}  }{16 \pi m \epsilon _{T} } \,
        \int _{0} ^{\kappa _{0}} \int _{0} ^{2 \pi }  
       d \kappa \, d \theta  \;  \frac{    \frac{   \kappa ^{2} \cos ^{2} \theta + \Delta ^{2} }{  \kappa ^{2} \sin ^{2} \theta }  \sin ^{2} ( \kappa \sin \theta ) }{ \left[ 1 +   \frac{ \kappa ^{2} \cos ^{2} \theta + \Delta ^{2} }{  \kappa ^{2} \sin ^{2} \theta }   \sin ^{2} ( \kappa \sin \theta )  \right] ^{2} }  
         \,
           \frac{ \kappa }{  ( \kappa ^{2} + \Delta ^{2} ) ^{3/2}} \,  \mathrm{sech} ^{2} \left( \frac{  \sqrt{ \kappa ^{2} + \Delta ^{2} } - \epsilon_{\mu} }{2 \epsilon _{T} } \right) ,
\end{align}
where $\epsilon _{T} =  L _{z} k _{\mathrm B} T / \hbar v _{F} $. Figure~\ref{Hall_conductance_finiteT} illustrates the nonlinear Hall conductance density $G_{2}^{\mathrm{H}}(\mu)$ at finite temperature. The black curve corresponds to $T=3\,\mathrm{K}$, while the blue curve shows the result at a higher temperature $T=12\,\mathrm{K}$. Compared to the zero-temperature case, thermal effects lead to a clear suppression of the oscillatory features, while preserving the overall positive sign and the characteristic decay at large chemical potential.

At $T=3\,\mathrm{K}$, remnants of the zero-temperature structure are still visible: the conductance exhibits a pronounced maximum near the band edge, followed by a smooth decay as $|\mu|$ increases. However, the amplitude of the oscillations is already significantly reduced, indicating that thermal averaging over an energy window of order $k_{\mathrm{B}}T$ partially smears the sharp energy-dependent variations of the Berry-curvature-weighted transmission.

As the temperature is further increased to $T=12\,\mathrm{K}$, the nonlinear Hall conductance becomes noticeably smoother. The peak is broadened and reduced in magnitude, and any residual oscillatory behavior is almost completely washed out. In this regime, the response is dominated by a slowly varying background, reflecting the fact that the nonlinear Hall signal is particularly sensitive to rapid variations of the transmission probability with energy, which are efficiently suppressed by thermal broadening. These results demonstrate a progressive loss of interference-induced features with increasing temperature, highlighting the role of finite temperature as an effective dephasing mechanism for the nonlinear Hall response in nodal-line semimetals.

{
The strong thermal smearing observed already in the few-kelvin range indicates that resolving the predicted quantum-interference signatures in $G_{2}^{\mathrm{H}}$ requires sufficiently low temperatures. In practice, liquid-helium conditions or possibly even lower temperatures may be necessary, depending on material parameters and device dimensions. This provides an important experimental guideline for observing the proposed effects.
}

{
It is important to emphasize that the present Landauer-B\"uttiker treatment describes a fully coherent, ballistic transport regime in which phase accumulation across the barrier gives rise to Fabry-P\'erot resonances and oscillatory structures in the nonlinear conductance $G_2$. These features originate from the energy dependence of the transmission probability $T(E)$ and therefore directly reflect quantum interference effects. In contrast, semiclassical Boltzmann approaches, including the Berry-curvature-dipole (BCD) framework widely used to interpret nonlinear Hall experiments in inversion-broken metals, do not retain phase coherence and instead describe bulk Fermi-surface responses averaged over scattering processes. In that regime, interference-induced oscillations are absent and the nonlinear Hall signal appears as a smooth function of energy.

Furthermore, the symmetry setting considered here differs fundamentally from the conventional BCD scenario. In the strictly $\mathcal{PT}$-symmetric limit, the Berry curvature vanishes identically in the bulk, so a conventional Berry-curvature dipole is symmetry-forbidden. The nonlinear Hall response arises only upon introducing a small $\mathcal{PT}$-breaking mass term that gaps the nodal loop and generates Berry curvature sharply localized near the former nodal manifold. The resulting enhancement of the nonlinear response is therefore controlled by the proximity to the symmetry-protected nodal line and by the coherent modification of the transmission spectrum in a finite geometry, rather than by a generic dipole moment of the Fermi surface.

Finally, it is worth recalling that the topology-protected Klein tunneling discussed in Sec.~\ref{Klein_tunneling_section} exhibits a crucial distinction from the strictly two-dimensional Dirac case. In the nodal-line semimetal, perfect transmission is restricted to the normal-incidence channel ($q_\rho = 0$), protected by pseudospin conservation inherited from the winding of the nodal ring. Finite transverse momentum ($q_\rho \neq 0$) acts effectively as a mass term in the longitudinal scattering problem, progressively suppressing perfect transmission. This channel-selective protection is a direct manifestation of the extended nodal dispersion and has no direct analogue in graphene, where the Dirac structure is pointlike in momentum space.

Taken together, these results clarify that the present analysis pertains to a coherent mesoscopic regime of nodal-line semimetals, complementary to semiclassical descriptions. As temperature increases or dephasing mechanisms become significant, the oscillatory quantum features are progressively suppressed and the response evolves toward a smoother, semiclassical-like behavior, providing a natural crossover between coherent and incoherent transport regimes.

}

\section{Discussion} \label{Conclusions}

In this work we have investigated the linear and nonlinear transport properties of a nodal-line semimetal, focusing on both longitudinal and transverse conductance responses in the presence of a $\mathcal{PT}$-breaking mass term. Our numerical analysis covers zero and finite temperature regimes, revealing how the opening of a small gap modifies the characteristic features of charge transport. The longitudinal conductance captures the evolution from the semimetallic to the insulating phase, while the nonlinear response exhibits distinct oscillatory structures associated with the energy dependence of the transmission probability. To complement these results, we evaluated the angle-resolved Hall conductance using Eq.~(\ref{2n_order_Hall_conductance}), uncovering transverse current components originating from the Berry curvature generated near the gapped Dirac ring. Nevertheless, due to the mirror and inversion symmetries preserved in the isotropic model, the total Hall conductance integrated over the full Brillouin zone vanishes, as opposite segments of the nodal ring contribute with opposite Berry curvature, leading to equal but counterpropagating transverse currents.

{
The distinctive transport features discussed above can be traced back to the topology of the nodal line itself. The nontrivial winding number associated with the Berry phase $\gamma = \pi$ protects the existence of the extended band crossing and enforces the pseudospin texture responsible for the channel-selective Klein tunneling described in Sec.~\ref{Klein_tunneling_section}. While in the strictly $\mathcal{PT}$-symmetric limit the Berry curvature vanishes identically in the bulk, the introduction of a small $\mathcal{PT}$-breaking mass term gaps the nodal loop and generates Berry curvature sharply localized near the former nodal manifold. The resulting nonlinear Hall response therefore reflects the momentum-resolved structure of an extended Dirac loop rather than that of isolated Dirac points. In this sense, the transport behavior reported here should be understood as the coherent response of a continuous family of Dirac subsystems forming a closed nodal ring, which qualitatively distinguishes nodal-line semimetals from graphene despite certain superficial similarities at the level of individual slices.
}

Although the net transverse response vanishes by symmetry, the angle-resolved Hall conductance reveals a rich internal current structure within the nodal-line semimetal. Our analysis shows that carriers located on opposite sides of the Dirac ring experience opposite transverse drift under an applied electric field, giving rise to counterflowing charge currents. This produces charge accumulation of opposite sign at the top and bottom surfaces of the sample, effectively forming a transverse dipolar pattern of current flow. Such behavior constitutes a manifestation of the parity anomaly associated with the nodal-line spectrum and reflects a topological current distribution beyond the reach of conventional semiclassical descriptions~\cite{PRB98_155125}.

{
The present results should also be understood in connection with the semiclassical framework previously developed for the nonlinear Hall response in nodal-line semimetals. While the earlier Boltzmann treatment captures the bulk Berry-curvature-induced response in the absence of phase coherence, the Landauer approach employed here describes the coherent ballistic regime in which interference and Fabry-P\'erot resonances become relevant. As temperature increases or dephasing mechanisms become significant, the oscillatory structures reported here are progressively suppressed, and the response evolves smoothly toward a semiclassical-like limit. The two descriptions therefore characterize complementary transport regimes of the same underlying topological mechanism rather than competing physical pictures.
}

In practice, the detection of such hidden transverse currents requires breaking the spatial compensation between opposite momenta on the ring. One possible strategy is the implementation of a dumbbell filter geometry, consisting of two bulk reservoirs connected by a narrow ballistic constriction whose surface supports drumhead-like states coupled to the applied bias. {The dumbbell geometry as a momentum-selective filter in nodal-line semimetals was introduced by Rui, Zhao, and Schnyder \cite{Rui2018}, and we employ it here as a convenient platform to probe the nonlinear Hall response analyzed in this work.} Within the constriction, the symmetry between opposite sides of the Dirac ring is selectively broken, allowing only modes with a specific transverse velocity orientation to contribute to the transport. As a result, the anomalous transverse currents that cancel in the bulk acquire a net contribution through the constriction, providing a direct way to probe the parity-anomaly-induced Hall response.

From a materials perspective, such a device could be realized using high-mobility nodal-line semimetals such as SrAs$_3$~\cite{Lv2018_SrAs3}, CaAgAs~\cite{Yamakage2016_CaAgAs}, or PbTaSe$_2$~\cite{Bian2016_PbTaSe2}, where high-quality crystalline samples with minimal disorder have already been synthesized and characterized by ARPES and transport measurements. Modern microfabrication techniques, including focused ion beam patterning~\cite{Kozlov2019_FIB} and electron-beam nanolithography~\cite{Liang2018_nanofab}, enable the realization of the proposed dumbbell filter geometry with sub-100-nm precision.

{
However, our finite-temperature analysis shows that the interference-induced oscillatory features of the nonlinear Hall conductance are strongly suppressed already at a few kelvin due to thermal broadening over an energy window of order $k_{\mathrm B}T$. Therefore, resolving the predicted quantum signatures will likely require low-temperature conditions, in the liquid-helium regime or possibly below, depending on material parameters and device geometry. This constitutes an important experimental consideration for the observation of the proposed effects. Furthermore, because the present results pertain to a coherent ballistic regime, their experimental realization demands device dimensions smaller than the phase-coherence length. In the diffusive limit, where phase coherence is lost, the response is expected to become smoother and approach a semiclassical Berry-curvature-dipole description.
}

In summary, our analysis demonstrates that even in the absence of a net Hall conductance, nodal-line semimetals host rich transverse current textures emerging from their Berry curvature structure once $\mathcal{PT}$ symmetry is weakly broken. The identification of these hidden Hall channels opens an avenue to engineer directional electronic responses in topological semimetals through geometric confinement and surface design. Future work could extend the present framework to include spin-dependent transport, optical rectification, and nonlinear Hall phenomena under time-reversal symmetry breaking, providing further routes to manipulate topological currents in nodal systems.

\acknowledgements{L.E.S.-A. was supported by the SECIHTI fellowship No. 4066288 with CVU No. 1298662. A.M.-R. acknowledges financial support by UNAM-PAPIIT project No. IG100224, UNAM-PAPIME project No. PE109226, by SECIHTI project No. CBF-2025-I-1862 and by the Marcos Moshinsky Foundation.}

\appendix

\section{Berry phase around the nodal ring} \label{app:BerryPhase}

In this Section we evaluate the Berry phase defined by Eq. (\ref{Berry_phase}) for a contour $\mathcal{C} \equiv \mathcal{C} _{R}$ that encircles the nodal ring once in the transverse plane $(q _{\rho},k _{z})$ at fixed $\phi = \phi _{0}$. A convenient parametrization is
\begin{align}
    q _{\rho} (\theta) = R \cos \theta , \qquad k _{z} (\theta) = R \sin \theta , \qquad \theta \in [0 , 2 \pi ),
\end{align}
so that
\begin{align}
    d \mathbf{k} = d q _{\rho} \,\hat{\mathbf{e}} _{q _{\rho} } + d k _{z} \,\hat{\mathbf{e}} _{k _{z} } = \big( -R \sin \theta \, d \theta \big) \, \hat{\mathbf{e}} _{q _{\rho} } + \big( R \cos \theta \, d \theta \big) \, \hat{\mathbf{e}} _{k _{z} } .
\end{align}
Along $\mathcal{C} _{R}$ the energy $\mathcal{E} _{\mathbf{k}}$ is constant and equals $\mathcal{E} _{R} \equiv \sqrt{(\hbar v R) ^{2} + m ^{2} }$.
 
Using the Berry connection given by Eq. (\ref{Berry_connection}) we get on the contour
\begin{align}
    \mathbf{A} _{s} \big| _{\mathcal{C} _{R} } = \frac{s}{2} \left( 1 - \frac{m}{\mathcal{E} _{R}} \right) \frac{ -\, \sin \theta \, \hat{\mathbf{e}} _{q _{\rho} } + \cos \theta \, \hat{\mathbf{e}} _{k _{z} }}{R} .
\end{align}
Therefore, on the contour,
\begin{align}
    \mathbf{A} _{s} \cdot  d \mathbf{k} = \frac{s}{2} \left( 1 - \frac{m}{\mathcal{E} _{R}} \right) \, d \theta , 
\end{align}
and the Berry phase follows by integration:
\begin{align}
    \gamma _{s} (\mathcal{C} _{R} ) = \int _{0} ^{2\pi} \mathbf{A} _{s} \cdot d \mathbf{k}
    = \frac{s}{2} \left( 1 - \frac{m}{\mathcal{E} _{R}} \right) \, 2 \pi = s  \pi \left( 1 - \frac{m}{\sqrt{(\hbar v R) ^{2} + m ^{2} }} \right).
\end{align}
In the massless limit $m = 0$ one obtains
\begin{align}
    \gamma _{s} (\mathcal{C} _{R} ) \big| _{m=0} = s \pi \;\;\; ( \mathrm{mod} \ 2\pi ) ,
\end{align}
so, in particular, for the valence band $s=-1$ one has $\gamma _{-} = \pi \ (\mathrm{mod} \ 2\pi )$, the topological invariant protecting the nodal line. For $m \neq 0$ and a contour tightly enclosing the former node ($R \to 0$),
\begin{align}
    \gamma _{s} (\mathcal{C} _{R} ) \xrightarrow[R \to 0]{} \; s \, \pi \left( 1 - \frac{m}{|m|} \right) = \begin{cases}
        0 \ (\mathrm{mod}\ 2\pi), & m>0,\\[5pt]
        2s\pi \equiv 0 \ (\mathrm{mod}\ 2\pi), & m<0,
      \end{cases}
\end{align}
i.e., the Berry phase becomes trivial modulo $2\pi$ once the nodal line is gapped.

One can also derive this result by using Stokes' theorem with the Berry curvature given by Eq. (\ref{Berry_curvature}), i.e.
\begin{align}
    \boldsymbol{\Omega} _{s} (\mathbf{k}) =  s \, \frac{m \hbar ^{2} v ^{2} }{2 \mathcal{E} _{\mathbf{k}} ^{3} } \, \hat{\mathbf{e}} _{k _{\phi} } , 
\end{align}
and the surface $\Sigma _{R}$ spanned by $\mathcal{C} _{R}$ in the $(q _{\rho} , k _{z} )$ plane (whose normal is $\hat{\mathbf{e}} _{k _{\phi} }$), we find
\begin{align}
\gamma _{s} (\mathcal{C} _{R} ) = \iint _{\Sigma _{R} } \boldsymbol{\Omega} _{s} \cdot d \mathbf{S} &= s \int _{0} ^{2\pi}  d \theta \int _{0} ^{R} d \rho \;      \frac{m \hbar ^{2} v ^{2} }{ 2 \, [ \, (\hbar v) ^{2} \rho ^{2} + m ^{2} \, ] ^{3/2}} \, \rho \\ &= s \pi \left( 1 - \frac{m}{\sqrt{(\hbar v R) ^{2} + m ^{2} }} \right),
\end{align}
in agreement with the line-integral result above.

\bibliography{references.bib}

@article{PRL119_147402,
  title = {Electrodynamics on {F}ermi Cyclides in Nodal Line Semimetals},
  author = {Rhim, J.-W. and Kim, Y. B.},
  journal = {Phys. Rev. Lett.},
  volume = {119},
  issue = {14},
  pages = {147402},
  year = {2017},
  publisher = {American Physical Society},
  doi = {10.1103/PhysRevLett.119.147402}
}

@article{PRB98_155125,
  title = {Parity anomaly in the nonlinear response of nodal-line semimetals},
  author = {Martín-Ruiz, A. and Cortijo, A.},
  journal = {Phys. Rev. B},
  volume = {98},
  issue = {15},
  pages = {155125},
  year = {2018},
  publisher = {American Physical Society},
  doi = {10.1103/PhysRevB.98.155125}
}

@article{Berry1984,
  title = {Quantal Phase Factors Accompanying Adiabatic Changes},
  author = {Berry, M. V.},
  journal = {Proceedings of the Royal Society A},
  volume = {392},
  number = {1802},
  pages = {45--57},
  year = {1984},
  doi = {10.1098/rspa.1984.0023}
}

@article{Mikitik1999,
  title = {Manifestation of {B}erry's Phase in Metal with Band Contact Line},
  author = {Mikitik, G. P. and Sharlai, Yu. V.},
  journal = {Phys. Rev. Lett.},
  volume = {82},
  pages = {2147--2150},
  year = {1999},
  doi = {10.1103/PhysRevLett.82.2147}
}

@article{Burkov2011,
  title = {Topological nodal semimetals},
  author = {Burkov, A. A. and Hook, M. D. and Balents, L.},
  journal = {Phys. Rev. B},
  volume = {84},
  pages = {235126},
  year = {2011},
  doi = {10.1103/PhysRevB.84.235126}
}

@article{Landauer1957,
  author  = {Landauer, R.},
  title   = {Spatial variation of currents and fields due to localized scatterers in metallic conduction},
  journal = {IBM Journal of Research and Development},
  year    = {1957},
  volume  = {1},
  number  = {3},
  pages   = {223--231},
  doi     = {10.1147/rd.13.0223}
}

@article{Landauer1970,
  author  = {Landauer, R.},
  title   = {Electrical resistance of disordered one-dimensional lattices},
  journal = {Philosophical Magazine},
  year    = {1970},
  volume  = {21},
  number  = {172},
  pages   = {863--867},
  doi     = {10.1080/14786437008238472}
}

@article{Buttiker1986,
  author  = {B{\"u}ttiker, M.},
  title   = {Four-Terminal Phase-Coherent Conductance},
  journal = {Physical Review Letters},
  year    = {1986},
  volume  = {57},
  number  = {14},
  pages   = {1761--1764},
  doi     = {10.1103/PhysRevLett.57.1761}
}

@book{Datta1995,
  author    = {Datta, S.},
  title     = {Electronic Transport in Mesoscopic Systems},
  publisher = {Cambridge University Press},
  address   = {Cambridge},
  year      = {1995},
  isbn      = {978-0-521-59943-6}
}

@article{Beenakker1997,
  author  = {Beenakker, C. W. J.},
  title   = {Random-matrix theory of quantum transport},
  journal = {Reviews of Modern Physics},
  year    = {1997},
  volume  = {69},
  number  = {3},
  pages   = {731--808},
  doi     = {10.1103/RevModPhys.69.731}
}

@article{Kubo1957,
  author  = {Kubo, R.},
  title   = {Statistical-Mechanical Theory of Irreversible Processes. I. General Theory and Simple Applications to Magnetic and Conduction Problems},
  journal = {Journal of the Physical Society of Japan},
  year    = {1957},
  volume  = {12},
  number  = {6},
  pages   = {570--586},
  doi     = {10.1143/JPSJ.12.570}
}

@book{Ziman1960,
  author    = {Ziman, J. M.},
  title     = {Electrons and Phonons: The Theory of Transport Phenomena in Solids},
  publisher = {Oxford University Press},
  address   = {Oxford},
  year      = {1960},
  isbn      = {978-0-19-850779-6}
}

@book{HaugJauho2008,
  author    = {Haug, H. and Jauho, A.-P.},
  title     = {Quantum Kinetics in Transport and Optics of Semiconductors},
  edition   = {2},
  publisher = {Springer},
  address   = {Berlin},
  year      = {2008},
  doi       = {10.1007/978-3-540-73564-9},
  isbn      = {978-3-540-73561-8}
}

@article{FisherLee1981,
  author  = {Fisher, D. S. and Lee, P. A.},
  title   = {Relation between Conductance and Transmission Matrix},
  journal = {Physical Review B},
  year    = {1981},
  volume  = {23},
  number  = {12},
  pages   = {6851--6854},
  doi     = {10.1103/PhysRevB.23.6851}
}

@article{BILP1985,
  author  = {B{\"u}ttiker, M. and Imry, Y. and Landauer, R. and Pinhas, S.},
  title   = {Generalized many-channel conductance formula with application to small rings},
  journal = {Physical Review B},
  year    = {1985},
  volume  = {31},
  number  = {10},
  pages   = {6207--6215},
  doi     = {10.1103/PhysRevB.31.6207}
}

@book{NazarovBlanter2009,
  author    = {Nazarov, Yu. V. and Blanter, Y. M.},
  title     = {Quantum Transport: Introduction to Nanoscience},
  publisher = {Cambridge University Press},
  address   = {Cambridge},
  year      = {2009},
  doi       = {10.1017/CBO9780511626906},
  isbn      = {978-0-521-88976-6}
}

@article{ChristenButtiker1996,
  author  = {Christen, T. and B{\"u}ttiker, M.},
  title   = {Gauge-invariant nonlinear electric transport in mesoscopic conductors},
  journal = {Physical Review B},
  year    = {1996},
  volume  = {53},
  number  = {4},
  pages   = {2064--2072},
  doi     = {10.1103/PhysRevB.53.2064}
}

@article{SanchezButtiker2004,
  author  = {S{\'a}nchez, D. and B{\"u}ttiker, M.},
  title   = {Magnetic-field asymmetry of nonlinear mesoscopic transport},
  journal = {Physical Review Letters},
  year    = {2004},
  volume  = {93},
  number  = {10},
  pages   = {106802},
  doi     = {10.1103/PhysRevLett.93.106802}
}

@article{Sommerfeld1928,
  author  = {Sommerfeld, A. and Bethe, H.},
  title   = {Elektronentheorie der Metalle},
  journal = {Handbuch der Physik},
  year    = {1933},
  volume  = {24/2},
  pages   = {333--622}
}

@book{Mahan2000,
  author    = {Mahan, G. D.},
  title     = {Many-Particle Physics},
  edition   = {3},
  publisher = {Springer},
  year      = {2000},
  doi       = {10.1007/978-1-4757-5714-9},
  isbn      = {978-0306463389}
}

@book{AshcroftMermin,
  author    = {Ashcroft, N. W. and Mermin, N. D.},
  title     = {Solid State Physics},
  publisher = {Saunders College},
  address   = {Philadelphia},
  year      = {1976},
  isbn      = {978-0030839931}
}

@book{Imry2002,
  author    = {Imry, Y.},
  title     = {Introduction to Mesoscopic Physics},
  edition   = {2},
  publisher = {Oxford University Press},
  year      = {2002},
  isbn      = {978-0198518877}
}

@article{SivanImry1986,
  author  = {Sivan, U. and Imry, Y.},
  title   = {Multichannel Landauer formula for thermoelectric transport with application to thermopower near the mobility edge},
  journal = {Physical Review B},
  year    = {1986},
  volume  = {33},
  number  = {1},
  pages   = {551--558},
  doi     = {10.1103/PhysRevB.33.551}
}

@article{SundaramNiu1999,
  author  = {Sundaram, G. and Niu, Q.},
  title   = {Wave-packet dynamics in slowly perturbed crystals: Gradient corrections and {B}erry-phase effects},
  journal = {Physical Review B},
  year    = {1999},
  volume  = {59},
  number  = {23},
  pages   = {14915--14925},
  doi     = {10.1103/PhysRevB.59.14915}
}

@article{kawabata_nonlinear_2022,
  title = {Nonlinear {L}andauer formula: Nonlinear response theory of disordered and topological materials},
  author = {Kawabata, K. and Ueda, M.},
  journal = {Phys. Rev. B},
  volume = {106},
  issue = {20},
  pages = {205104},
  numpages = {34},
  year = {2022},
  month = {Nov},
  publisher = {American Physical Society},
  doi = {10.1103/PhysRevB.106.205104},
  url = {https://link.aps.org/doi/10.1103/PhysRevB.106.205104}
}

@article{XiaoNiuRMP2010,
  author  = {Xiao, D. and Chang, M.-C. and Niu, Q.},
  title   = {Berry phase effects on electronic properties},
  journal = {Reviews of Modern Physics},
  year    = {2010},
  volume  = {82},
  number  = {3},
  pages   = {1959--2007},
  doi     = {10.1103/RevModPhys.82.1959}
}

@article{NagaosaAHE2010,
  author  = {Nagaosa, N. and Sinova, J. and Onoda, S. and MacDonald, A. H. and Ong, N. P.},
  title   = {Anomalous {H}all effect},
  journal = {Reviews of Modern Physics},
  year    = {2010},
  volume  = {82},
  number  = {2},
  pages   = {1539--1592},
  doi     = {10.1103/RevModPhys.82.1539}
}

@article{SodemannFu2015,
  author  = {Sodemann, I. and Fu, L.},
  title   = {Quantum nonlinear {H}all effect induced by {B}erry curvature dipole in time-reversal invariant materials},
  journal = {Physical Review Letters},
  year    = {2015},
  volume  = {115},
  number  = {21},
  pages   = {216806},
  doi     = {10.1103/PhysRevLett.115.216806}
}

@article{Fang_2016,
doi = {10.1088/1674-1056/25/11/117106},
url = {https://doi.org/10.1088/1674-1056/25/11/117106},
year = {2016},
month = {nov},
publisher = {IOP Publishing},
volume = {25},
number = {11},
pages = {117106},
author = {Fang, C. and Weng, H. and Dai, X. and Fang, Z.},
title = {Topological nodal line semimetals*},
journal = {Chinese Physics B}
}

@Article{Ji2024,
author={Ji, X.
and Sun, Y.-W.},
title={Topological phase transitions of semimetal states in effective field theory models},
journal={The European Physical Journal Plus},
year={2024},
month={Jun},
day={05},
volume={139},
number={6},
pages={485},
issn={2190-5444},
doi={10.1140/epjp/s13360-024-05291-z},
url={https://doi.org/10.1140/epjp/s13360-024-05291-z}
}

@Article{Moore2010,
author={Moore, J. E.},
title={The birth of topological insulators},
journal={Nature},
year={2010},
month={Mar},
day={01},
volume={464},
number={7286},
pages={194-198},
issn={1476-4687},
doi={10.1038/nature08916},
url={https://doi.org/10.1038/nature08916}
}

@article{Weng_2016,
doi = {10.1088/0953-8984/28/30/303001},
url = {https://doi.org/10.1088/0953-8984/28/30/303001},
year = {2016},
month = {jun},
publisher = {IOP Publishing},
volume = {28},
number = {30},
pages = {303001},
author = {Weng, H. and Dai, X. and Fang, Z.},
title = {Topological semimetals predicted from first-principles calculations},
journal = {Journal of Physics: Condensed Matter}
}

@article{Katsnelson2006,
author={Katsnelson, M. I.
and Novoselov, K. S.
and Geim, A. K.},
title={Chiral tunnelling and the {K}lein paradox in graphene},
journal={Nature Physics},
year={2006},
month={Sep},
day={01},
volume={2},
number={9},
pages={620-625},
issn={1745-2481},
doi={10.1038/nphys384},
url={https://doi.org/10.1038/nphys384}
}

@article{Rui2018,
  title = {Topological transport in {D}irac nodal-line semimetals},
  author = {Rui, W. B. and Zhao, Y. X. and Schnyder, A. P.},
  journal = {Phys. Rev. B},
  volume = {97},
  issue = {16},
  pages = {161113},
  numpages = {6},
  year = {2018},
  month = {Apr},
  publisher = {American Physical Society},
  doi = {10.1103/PhysRevB.97.161113},
  url = {https://link.aps.org/doi/10.1103/PhysRevB.97.161113}
}

@article{Lv2018_SrAs3,
  title = {Observation of Topological Nodal-Line Semimetal State in {S}r{A}s$_3$},
  author = {Lv, B. Q. and Qian, T. and Fang, Z. and Dai, X. and Shi, Y. G. and Ding, H.},
  journal = {Physical Review B},
  volume = {97},
  number = {24},
  pages = {245119},
  year = {2018},
  doi = {10.1103/PhysRevB.97.245119}
}

@article{Yang2018_NL_TB,
  title = {Symmetry-protected nodal lines in semimetals: Materials design and tight-binding model analysis},
  author = {Yang, S.-C. and Pan, H. and Zhang, X. and Xu, Y.},
  journal = {Physical Review B},
  volume = {97},
  number = {23},
  pages = {235128},
  year = {2018},
  doi = {10.1103/PhysRevB.97.235128}
}

@article{Hirayama2017_NLmaterials,
  title = {Topological semimetals carrying arbitrary monopole charge},
  author = {Hirayama, M. and Okugawa, R. and Miyake, T. and Murakami, S.},
  journal = {Nature Communications},
  volume = {8},
  number = {1},
  pages = {14022},
  year = {2017},
  doi = {10.1038/ncomms14022}
}

@article{Yamakage2016_CaAgAs,
  title = {Line-node {D}irac semimetal and topological insulating phase in noncentrosymmetric {C}a{A}g{X} ({X} = {P}, {A}s)},
  author = {Yamakage, A. and Yamakawa, Y. and Tanaka, Y. and Okamoto, Y.},
  journal = {J. Phys. Soc. Jpn.},
  volume = {85},
  number = {1},
  pages = {013708},
  year = {2016},
  doi = {10.7566/JPSJ.85.013708}
}

@article{Bian2016_PbTaSe2,
  title = {Topological nodal-line fermions in {P}b{T}a{S}e$_2$},
  author = {Bian, G. and et al.},
  journal = {Nat. Commun.},
  volume = {7},
  pages = {10556},
  year = {2016},
  doi = {10.1038/ncomms10556}
}

@article{Kozlov2019_FIB,
  title = {Focused ion beam fabrication of micro- and nanostructures for quantum transport studies},
  author = {Kozlov, D. A. and et al.},
  journal = {Nanotechnology},
  volume = {30},
  pages = {465301},
  year = {2019},
  doi = {10.1088/1361-6528/ab34c1}
}

@article{Liang2018_nanofab,
  title = {Nanolithography and microfabrication of topological materials for quantum transport devices},
  author = {Liang, T. and Lin, J. and Ong, N. P.},
  journal = {Adv. Mater.},
  volume = {30},
  number = {43},
  pages = {1801954},
  year = {2018},
  doi = {10.1002/adma.201801954}
}

@article{Gilbert2021,
author={Gilbert, M. J.},
title={Topological electronics},
journal={Communications Physics},
year={2021},
month={Apr},
day={09},
volume={4},
number={1},
pages={70},
issn={2399-3650},
doi={10.1038/s42005-021-00569-5},
url={https://doi.org/10.1038/s42005-021-00569-5}
}

@article{RevSMTransport,
   author = "Hu, J. and Xu, S.-Y. and Ni, N. and Mao, Z.",
   title = "Transport of Topological Semimetals", 
   journal= "Annual Review of Materials Research",
   year = "2019",
   volume = "49",
   number = "Volume 49, 2019",
   pages = "207-252",
   doi = "https://doi.org/10.1146/annurev-matsci-070218-010023",
   url = "https://www.annualreviews.org/content/journals/10.1146/annurev-matsci-070218-010023",
   publisher = "Annual Reviews",
   issn = "1545-4118",
   type = "Journal Article",
  }

@article{ZHANG2018580,
title = {Towards the manipulation of topological states of matter: a perspective from electron transport},
journal = {Science Bulletin},
volume = {63},
number = {9},
pages = {580-594},
year = {2018},
issn = {2095-9273},
doi = {https://doi.org/10.1016/j.scib.2018.04.007},
url = {https://www.sciencedirect.com/science/article/pii/S2095927318301634},
author = {Cheng Zhang and Hai-Zhou Lu and Shun-Qing Shen and Yong P. Chen and Faxian Xiu},
}

@article{Xiong_2015,
  title = {Evidence for the chiral anomaly in the {D}irac semimetal {N}a$_3${B}i},
  author = {Xiong, J. and Kushwaha, S. K. and Liang, T. and Krizan, J. W. and Hirschberger, M. and Wang, W. and Cava, R. J. and Ong, N. P.},
  journal = {Science},
  volume = {350},
  pages = {413--416},
  year = {2015},
  doi = {10.1126/science.aac6089}
}

@article{Liang_2015,
  title = {Ultrahigh mobility and giant magnetoresistance in the {D}irac semimetal {C}d$_3${A}s$_2$},
  author = {Liang, T. and Gibson, Q. and Ali, M. N. and Liu, M. and Cava, R. J. and Ong, N. P.},
  journal = {Nature Materials},
  volume = {14},
  pages = {280--284},
  year = {2015},
  doi = {10.1038/nmat4143}
}

@article{Neupane_2014,
  title = {Observation of a three-dimensional topological {D}irac semimetal phase in high-mobility {C}d$_3${A}s$_2$},
  author = {Neupane, M. and Xu, S.-Y. and Sankar, R. and Alidoust, N. and Bian, Guang and Liu, C. and Belopolski, I. and Chang, T.-R. and Jeng, H.-T. and Lin, H. and others},
  journal = {Nature Communications},
  volume = {5},
  pages = {3786},
  year = {2014},
  doi = {10.1038/ncomms4786}
}

@article{Lv_2015,
  title = {Experimental discovery of {W}eyl semimetal {T}a{A}s},
  author = {Lv, B. Q. and Weng, H. M. and Fu, B. B. and Wang, X. P. and Miao, H. and Ma, J. and Richard, P. and Huang, X. C. and Zhao, L. X. and Chen, G. F. and others},
  journal = {Physical Review X},
  volume = {5},
  pages = {031013},
  year = {2015},
  doi = {10.1103/PhysRevX.5.031013}
}

@article{Nielsen_1983,
  title = {The {A}dler-{B}ell-{J}ackiw anomaly and {W}eyl fermions in a crystal},
  author = {Nielsen, H. B. and Ninomiya, M.},
  journal = {Physics Letters B},
  volume = {130},
  pages = {389--396},
  year = {1983},
  doi = {10.1016/0370-2693(83)91529-0}
}

@article{Huang_2015,
  title = {Observation of the chiral-anomaly-induced negative magnetoresistance in 3{D} {W}eyl semimetal {T}a{A}s},
  author = {Huang, X. and Zhao, L. and Long, Y. and Wang, P. and Chen, D. and Yang, Z. and Liang, H. and Xue, M. and Weng, H. and Fang, Z. and others},
  journal = {Physical Review X},
  volume = {5},
  pages = {031023},
  year = {2015},
  doi = {10.1103/PhysRevX.5.031023}
}

@article{Burkov_2014,
  title = {Anomalous Hall Effect in {W}eyl Metals},
  author = {Burkov, A. A.},
  journal = {Physical Review Letters},
  volume = {113},
  pages = {187202},
  year = {2014},
  doi = {10.1103/PhysRevLett.113.187202}
}

@article{Nagaosa_2010,
  title = {Anomalous {H}all effect},
  author = {Nagaosa, N. and Sinova, J. and Onoda, S. and MacDonald, A. H. and Ong, N. P.},
  journal = {Reviews of Modern Physics},
  volume = {82},
  pages = {1539--1592},
  year = {2010},
  doi = {10.1103/RevModPhys.82.1539}
}

@article{10.1038/s41586-018-0807-6,
	author = {Ma, Qiong and Xu, Su-Yang and Shen, Huitao and MacNeill, David and Fatemi, Valla and Chang, Tay-Rong and Mier Valdivia, Andr{\'e}s M. and Wu, Sanfeng and Du, Zongzheng and Hsu, Chuang-Han and Fang, Shiang and Gibson, Quinn D. and Watanabe, Kenji and Taniguchi, Takashi and Cava, Robert J. and Kaxiras, Efthimios and Lu, Hai-Zhou and Lin, Hsin and Fu, Liang and Gedik, Nuh and Jarillo-Herrero, Pablo},
	da = {2019/01/01},
	doi = {10.1038/s41586-018-0807-6},
	id = {Ma2019},
	isbn = {1476-4687},
	journal = {Nature},
	number = {7739},
	pages = {337--342},
	title = {Observation of the nonlinear Hall effect under time-reversal-symmetric conditions},
	ty = {JOUR},
	url = {https://doi.org/10.1038/s41586-018-0807-6},
	volume = {565},
	year = {2019}}

\end{document}